# Mega Influencers versus Niche Creators:

# An Empirical Study of Streamer Influence on Endorsed Product Usage

Wooyong Jo, Mike Lewis, Yanwen Wang

***Abstract***

Social media has given rise to online consumption communities, or fandoms, which are complex networks of ancillary creators and consumers organized around a core product or intellectual property. Video game communities, for example, link players with content creators centered on a specific game. These networks are strategically complex: publishers often sponsor creators, yet the two can have divergent incentives, as creators may benefit from content that grows their own following at the core game's expense. We investigate the causal effect of consuming live-streamed content on subsequent gameplay for a specific game, exploiting an unexpected service interruption of the livestreaming platform together with time zone differences among users. Live-streamed content significantly increases gameplay: a 10% increase in live-streamed content viewing minutes yields a 3.43% increase in gameplay minutes. We further examine how this effect varies with how users distribute their viewing across streamer types, namely mega streamers with massive audiences, micro streamers offering easier interactivity, and publisher channels emphasizing official content. The positive effects are strongest when viewing skews toward micro streamers and weakest when it skews toward mega streamers. These patterns replicate across a broad set of other video games, supporting external validity. The findings inform how firms allocate sponsorship resources across creator tiers.

---

Wooyong Jo (wooyong@purdue.edu) is Assistant Professor of Marketing at Daniels School of Business at Purdue University, Mike Lewis (mike.lewis@emory.edu) is Professor of Marketing at the Goizueta School of Business, Emory University, Yanwen Wang (yanwen.wang@sauder.ubc.ca) is Associate Professor of Marketing at the Sauder School of Business, University of British Columbia.

## 1. Introduction

Livestreaming through online platforms like Twitch, YouTube, and Facebook has become a popular and impactful form of entertainment. The popularity is illustrated by a comparison of Twitch and Netflix. In 2024, Twitch hosted approximately 240 million unique visitors every month (Dean 2024), which was roughly ten times the number of Netflix's monthly active users (Spangler 2024). These massive audiences have been recognized by marketers, and marketing spending on streamers and other social media influencers has grown rapidly. Spending on creators on streaming and social media platforms was estimated to exceed $32.55 billion in 2025 (Influencer Marketing Hub 2026). Furthermore, survey data suggests that 84% of consumers have tried a product recommended by social media creators (Inmar Intelligence 2021). The academic literature on social media marketing and livestreaming is expanding (Hughes et al. 2019, Leung et al. 2022, Huang and Morozov 2025, Li et al. 2025), but there remains a lack of studies employing individual-level data explicitly linking exposure to streamer content to the consumer's subsequent product usage.

A complicating factor in researching livestreaming is the significant heterogeneity within the category. Livestreaming encompasses a variety of formats and genres, including live podcasts, musical performances, and video game matches. Our research is set in the video game category for several reasons. First, video games are a particularly popular source of live-streamed content. In January 2024, consumers watched over 1.65 billion hours of live-streamed content on Twitch, with the video game category accounting for about 80% of these hours (SullyGnome 2024).

Second, the video game streaming category also includes significant variation in the types of streamers. In addition to the vast number of people who use social media as a casual activity, star creators have emerged. For example, Tyler Blevins, a.k.a. Ninja, has amassed over 19 million followers on Twitch (Muller 2024) and Félix Lengyel, a.k.a. xQc, signed a non-exclusive contract with Kick, a livestreaming platform, worth up to $100 million over two years (Tassi 2023). The diversity in streamer types and audience sizes introduces additional dimensions to be considered when assessing streamer influence, as streamers like Ninja or xQc are themselves entertainment brands with their own goals and objectives.

Third, video game streaming is especially salient because of its role within gaming communities. In video game streaming, when an expert or charismatic player live broadcasts gameplay, it is not just popular entertainment but also a vital part of video game and esports communities. Livestreaming often spotlights a specific game and creates connections between players, building fandom communities (Jo and Lewis 2024). Many creators may focus on a particular game and play a crucial role in their respective gaming community by providing entertainment, expertise, and excitement (Jo and Lewis 2024, Huang and Morozov 2025, Li et al. 2025). The position of streamers in video game communities is critical because they explicitly promote the game. The role of streamers is analogous to that of ancillary content in other categories, such as commentary on sports or movies, where third parties enrich the consumption of cultural products.

Video games are also useful as they offer a clean empirical setting for studying the relationship between social media celebrities featuring products and consumer choices. Significant questions about the influence and impact of creators on video game consumers remain unanswered. For instance, the integral role of streamers within gaming communities has generated strong interest among publishers in sponsoring creators to grow and engage their player bases, making the question of how streamer content actually affects gameplay consequential. A fundamental question is whether streamers inspire additional gameplay or compete with games. Creators' content might stimulate more gameplay and increase fandom for the game, or creators may develop their own fan bases and compete with the game for consumers' attention. In addition to whether streamers provide complements to or substitutes for the game, there are questions about whether potential effects vary by viewer habits and streamer type. As noted, some streamers perform in front of massive audiences while others provide a more intimate experience.

Interactivity and repeat viewing are likely to have profound effects on creator influence, as the viewing experience becomes more relationship-oriented. A viewer who tends to watch streamers with massive audiences experiences a far less interactive experience than viewers who follow and repeatedly watch streamers with smaller audiences that allow for greater interaction. Intuitively, having personal exchanges with, and repeated viewing of, specific creators both contribute to the formation of a relationship between viewer and creator, as the relationship becomes more interactive and dynamic (Fournier 1998). Audience size may be a critical variable in this process because a creator with thousands of viewers cannot provide the same level of interactivity as a creator with a hundred viewers (Lu et al. 2021). However, the direction of an audience size effect is in question as a creator with a smaller, more interactive community might have stronger relationships and exert greater influence, while a creator with a massive audience might be viewed as more credible (Leung et al. 2022). Credibility might also be impacted by ownership and independence. Specifically, a publisher's channel may carry a different kind of credibility, as it serves as an official source for information such as updates, tournaments, or special events. However, some viewers may be skeptical due to its affiliation with the publisher.

Our research investigates the influence of creators within a specific gaming community. The game is one of the largest in the MOBA (Multiplayer Online Battle Arena) category, and our data includes information on user engagement with both the game and live streamers. The empirical challenge is to cleanly estimate the causal relationship between live-streamed content viewing and subsequent engagement with the focal game. To address this challenge, we leverage an unexpected outage of the streaming platform that did not affect the video game. The key to this natural experiment is that the outage differentially impacted users based on their local time zones. Our key identification comes from these differential exposures: users whose prime entertainment hours overlapped more with the Twitch outage experienced a larger reduction in streaming viewing availability than those whose prime hours did not. This cross-country and within-country variation in outage overlap generates vital exogenous differences in exposure to live-streamed content, which we use to identify the causal effect of livestreaming viewing on gameplay with a control function approach. We also replicate our core findings at the game-title level across hundreds of other titles with diverse characteristics, supporting the broader generalizability of our results.

Our research strategy involves developing a multilevel understanding of the role of creators. At the top level, we address the overarching question: Is live-streamed content a complement to or substitute for the source material (the game) within a gaming community? This is an essential question in the video game industry because most creators' content consists of shared screens of their gameplay. Our investigation of whether streamer content is a complement or substitute to gameplay supplements existing work on the topic by Huang and Morozov (2025). The use of individual-level data affords opportunities to evaluate more specific behaviors. In particular, we extend the top-level analysis to investigate more micro-level impacts of creators on individual players. First, we explore how creator content and gameplay are co-consumed. Using individual-level data, we examine whether live-streamed content is consumed concurrently with gameplay or in sequence. We then extend the analysis to examine how streamer audience characteristics and streamer repeat viewing affect creator influence.

Our analysis suggests that watching live-streamed content positively affects users' gameplay. The results indicate that livestreaming is a beneficial part of the gaming community, as streamer content drives incremental play rather than acting as a substitute. In terms of how gameplay and livestreaming are consumed, we find that sequential rather than concurrent consumption is more prevalent. We also find that the effects of livestreaming viewing are moderated by the types of streamers repeatedly viewed. Specifically, complementarity increases significantly for users whose repeated viewing is concentrated on micro streamers relative to mega streamers. We also find that a higher tendency to view the firm-owned channel yields an intermediate effect between the micro and mega streamer categories.

Our research primarily contributes to the literature on the creator economy and influencer marketing. It adds to the growing body of work examining how social media creator content influences consumer engagement. For example, Tian et al. (2024) and Cheng and Zhang (2025) examined issues related to video viewing and sponsorship on follower growth. Gu et al. (2024) and Feng et al. (2025) examined the effect of influencer characteristics on trust. Our study is most closely aligned with research on the connection between viewing live-streamed content and gameplay. The potential impact of content creators on game consumption highlights a critical aspect of gaming ecosystems: consumers interact within complex communities comprising games, players, and affiliated content creators. Success in the video game industry is often driven by the ability to create a vibrant, connected community in addition to creating a quality game.

Our work on creator-player relationships adds to the literature on video game ecosystem components and player engagement. Previous research has investigated several aspects of the relationship between livestreaming and gaming. Jo and Lewis (2024) examined the economics of publisher-produced live-streamed esports content and found that this content increased product usage and spending, largely through knowledge transfer effects. Huang and Morozov (2025) showed, using aggregate game-level data, that livestreaming generally complements gameplay, with stronger effects for smaller publishers, cheaper and highly rated titles, and niche games. Li et al. (2025) examined the impact of YouTube videos on gameplay and game title purchases. They found significant response heterogeneity as gameplay increased for adventure and sports games, but decreased for role-playing games. Purchases declined for story-driven, indie, and sports games, but rose for simulation and multiplayer games. Relatedly, Qian and Xie (2026) leveraged star creators' exits from Twitch and documented heterogeneous supply-side responses across creator types, with major creators reducing content provision more than micro creators. Whereas these studies examine heterogeneity across game genres or in creators' content supply, our analysis examines heterogeneity across creator types and the user-creator relationships, on the demand side, that aggregate data cannot recover.

In contrast to much of the previous research (Huang and Morozov 2025, Li et al. 2025), our study utilizes disaggregate data and focuses on "individual" player behavior and user-streamer relationships. Aggregate game-day data can establish whether viewing and play move together, but it cannot observe how an individual sequences the two activities, which streamers a given user repeatedly engages with, or how the same user's play responds to a viewing shock. Our linked individual-level data identify all three: we distinguish sequential from concurrent consumption, characterize user-streamer relationships across creator types, and estimate within-user viewing-to-play responses. This is what differentiates our contribution from aggregate studies such as Huang and Morozov (2025) and Li et al. (2025).

Our work is also notable because the individual-level playing and viewing data allows us to focus attention on a critical part of the fandom ecosystem: the relationships between consumers and ancillary content creators. Within the broader value-creation system of fandom communities, creators' engagement-building goals can be problematic if the content created directly competes with the core product. This is especially problematic when publishers consider sponsoring independent creators. Publishers wish to develop a broad, vibrant gaming community that includes independent creators, but there is an ever-present danger that viewers' concentrated attention on streamers does not always translate into engagement with the promoted game. Understanding the heterogeneous effects of different streaming environments requires observing how individual users distribute their repeated viewing across streamer types, which is central to our analysis.

## 2. Data and Constructs

Video games have evolved to have substantial appeal as games consumers play and as source material for entertainment for consumers to watch, such as live-streamed content and esports. While understanding the interplay of consumers' decisions to watch and/or play is crucial for game publishers, analyzing the video game ecosphere is challenging because data on gaming and watching are seldom linked. The data for our study is notable because it includes information on customer-level livestreaming viewing and gameplay. This section describes these datasets and provides operational definitions for our key variables.

### 2.1. Data

A major U.S. game publisher provided anonymized user-level data on gameplay and live-streamed content viewed on Twitch.tv for its most popular game title. This game falls into the MOBA genre, where players are placed onto two teams of five members and attempt to conquer the opposing team's base. The game publisher uses a freemium business model, meaning no upfront fee is required to play, and revenue is derived from optional purchases.

The sample of users was randomly drawn from nine distinct countries, with each country sample represented proportionally. The nine countries are among the game's top 12 markets and were selected to enable identification of the "outage" effect through time-zone differences. In addition to the U.S. (UTC-4; U.S. Eastern time), eight countries in Europe were selected: the U.K. (UTC+1), Spain (UTC+2), Germany (UTC+2), France (UTC+2), Poland (UTC+2), Romania (UTC+3), Ukraine (UTC+3), and Türkiye (UTC+3). Table 1 provides descriptive statistics of our sample by country.

Table 1: Samples by Country of Origin

| Country | Time Zone (Summer time) | Sample Size | Portion (%) | Cumulative Portion (%) |
|---|---|---|---|---|
| U.S. | UTC – 4h* | 6,150 | 60.69 | 60.69 |
| Spain | UTC + 2h | 1,209 | 11.93 | 72.62 |
| U.K. | UTC + 1h | 1,088 | 10.74 | 83.36 |
| Germany | UTC + 2h | 586 | 5.78 | 89.14 |
| France | UTC + 2h | 335 | 3.31 | 92.45 |
| Poland | UTC + 2h | 246 | 2.43 | 94.88 |
| Romania | UTC + 3h | 236 | 2.33 | 97.21 |
| Ukraine | UTC + 3h | 157 | 1.55 | 98.76 |
| Türkiye | UTC + 3h | 126 | 1.24 | 100.00 |
| Total | | 10,133 | 100.00 | |

Note: Given that the study period is during the summer, all UTC hours are based on daylight saving time. *As time zones vary by US state, we present Eastern daylight-saving time for simplicity.

The gameplay data includes player IDs, sign-up dates, countries of origin, and gameplay logs (i.e., timestamps of matches played). The livestreaming data includes viewing logs, which contain user identifiers, streamer identifiers, viewing timestamps (start and end), and video game identifiers being broadcast by the streamers. The viewing data was available to the firm because users voluntarily linked their Twitch profiles to their gaming accounts. Thus, our sample reflects players who chose to link their accounts across the two platforms, allowing for observations of both livestreaming and gaming.

The data spans nine weeks, from July 18, 2017, to September 18, 2017. The first four weeks of data are used as a calibration period to compute user-level characteristics. The remaining five weeks are used to estimate the effect of live-streamed content viewing on gameplay. These five weeks include an unexpected server outage of Twitch.tv, which we describe in detail in Section 3. Figure 1 shows the date ranges and time of the exogenous shock.

Figure 1: Observation Period

**7/18/2017**
**8/15/2017**
**Twitch Server Down Events**
(8/30/2017 – 8/31/2017)
**9/18/2017**
Calibration period (four weeks)
Two weeks before the shock week
The shock week
Two weeks after the shock week
*Time*
Main analysis period (five weeks)

Note: Color is available online.

2.2. Streamers and User–Streamer Ties

One of our goals is to understand the effect of watching different types of streamers on subsequent gameplay. We categorize the 5,424 streamer accounts based on audience size. The logic for our categorization is that audience size potentially impacts interactivity and credibility (Leung et al. 2022). The official channel is also considered a separate category, as it may be viewed differently by consumers. For instance, the official channel has access to inside information, but might also be perceived as a marketing tool.

We define streamers' popularity using the share of viewing minutes during the calibration period (first four weeks of data). Audience sizes on livestreaming platforms are highly skewed, with a small number of "mega" influencers and many micro influencers. We identify 36 major streamer accounts that together account for about 80% of game-related viewing minutes during this period. Of these, the single firm-owned channel is the most viewed, capturing about 18% of all viewing minutes, and the remaining 35 are independent mega streamers, collectively accounting for about 62%. We categorize the other 5,388 accounts, which together account for roughly 20% of viewing, as micro streamers. Figure 2 shows the distribution of viewership shares.

Figure 2: Distribution of Shares by Streamers

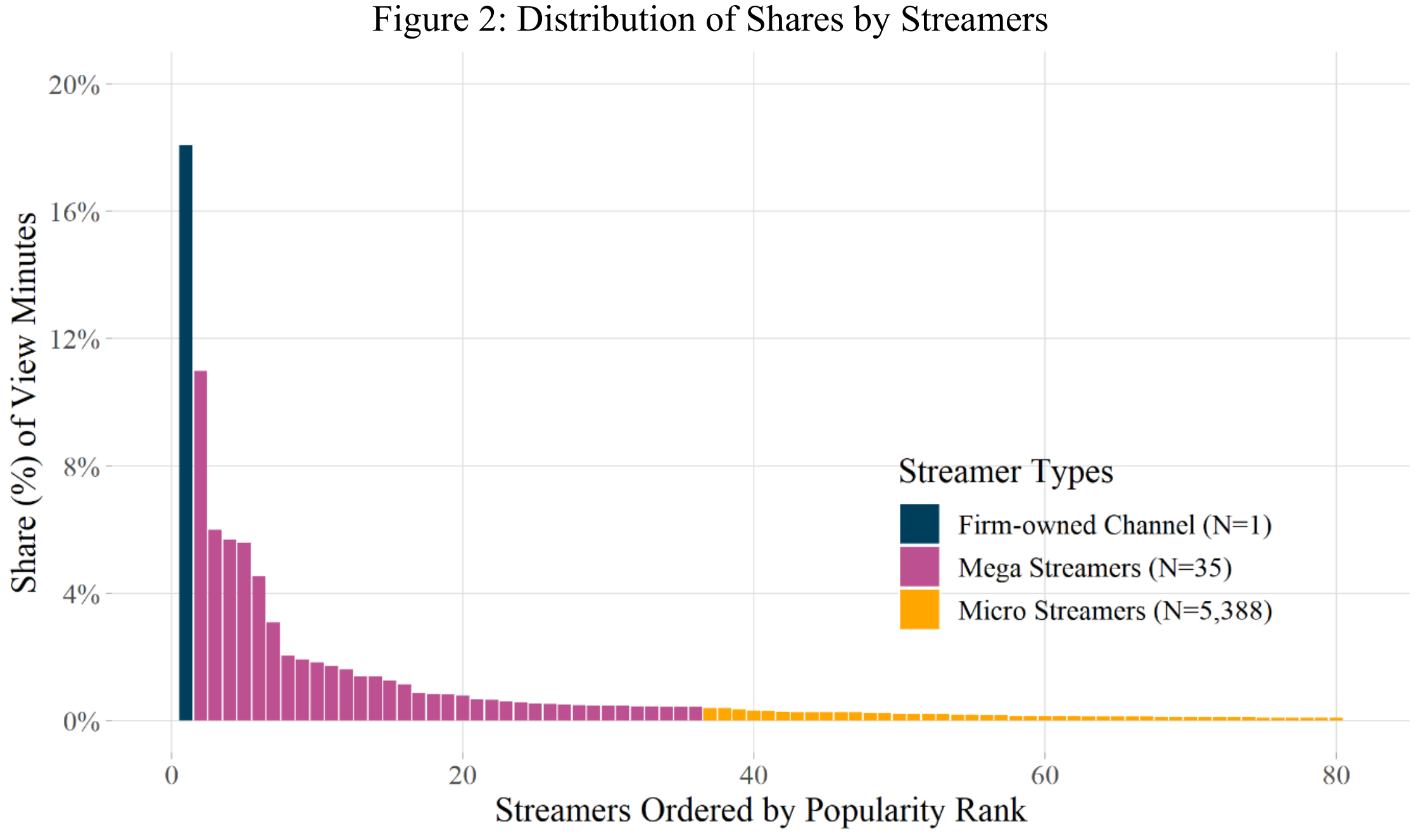

Note: We plot the shares of the top 80 streamers for display. Our sample has 5,424 streamers.

An important feature of our dataset is that it allows us to track individual players' repeated viewing of specific streamers over time, which provides insight into the role of user–streamer relationships in shaping influence. These data reveal not only the existence of user–streamer ties but also their strength, as measured by unique visits and viewing minutes. A complexity in social media viewing data is that much of the content offered to viewers is algorithmic, so the viewing of any single video or stream may be the product of a suggestion rather than a choice. The implication is that a single streamer view may result from a combination of user interests and algorithmic criteria. Repeat viewing of a streamer, however, provides a reliable signal that the user has a preference for the streamer. We begin our analysis of viewer-streamer relationships by evaluating repeat viewing metrics that speak to the strength of the connections.

Our goal is to construct measures that reveal content type preferences and streamer relationships. To construct these measures, we first need to define what constitutes a meaningful interaction as opposed to occasional sampling. Specifically, we categorize streamers as "meaningfully tied" if a user records ***eight*** or more unique visits during the four-week calibration period (about two visits per week). For robustness, we repeat the analysis using stricter (12 visits) and more relaxed (4 visits) thresholds.

Table 2: Distribution of Users with Meaningful Ties to Streamers

| *Meaningful Streamer Count* | *Users* | *Proportion of Streamers with Meaningful Ties* | | |
|---|---|---|---|---|
| | | *Firm-owned* | *Mega* | *Micro* |
| 0 | 4,969 | | | |
| 1 | 2,414 | .143 | .709 | .148 |
| 2 | 1,047 | .106 | .766 | .128 |
| 3 | 548 | .098 | .774 | .128 |
| 4 | 372 | .100 | .753 | .147 |
| 5+ | 783 | .071 | .739 | .189 |

Note: Sample includes 10,133 users. Viewing behaviors are based on the four-week calibration period. A streamer is classified as "meaningful" if a user records at least eight visits during this period.

Table 2 shows the prevalence of streamer-viewer ties. For instance, 4,969 viewers have no meaningful ties to any streamer, while 2,414 have a meaningful tie to a single streamer. For viewers with a single meaningful tie, 14.2% are meaningfully tied to the official channel, 70.9% to a mega streamer, and 14.8% to a micro streamer. We find that mega streamers consistently account for the largest share of meaningful ties (generally over 70%), highlighting their dominant role on the platform. For users prone to forming meaningful connections with streamers, the share of micro streamers increases slightly, indicating that users with broader engagement are more likely to connect with smaller creators.

As Table 2 shows, video game players form meaningful connections with only a small subset of the streamers they watch: on average, a user views about 11.2 streamers during the calibration period but maintains meaningful ties with only 1.3. This means that while users sample a broad range of streamers, the number of meaningful relationships is relatively limited. Among users who have at least one meaningful tie, more than 67% maintain such ties with only two or fewer streamers (3,461 out of 5,164). This concentration indicates that users' strongest connections are highly selective: although viewers can, in principle, engage repeatedly with many streamers, their repeated and sustained attention tends to center on just a few. Notably, micro streamers still account for a small share of meaningful ties (roughly 13 to 15%, Table 2). To provide additional context, we report in Web Appendix A how viewership of different streamer types varies across countries.

We collect data on meaningful ties because we are interested in the types of streamers that a viewer prefers. Our conjecture is that consumers who repeatedly view streamers with smaller (larger) audiences have different viewing experiences, either more (less) interactive. Ideally, we would investigate the role of loyalty to specific streamers. However, because of the large number of streamers, we focus on category-level viewership shares rather than ties to specific streamers, which we define in Section 4.5, where we examine how the effect of viewing varies with the types of streamers a user watches.

## 3. The Shock: Twitch Service Interruption

Challenges exist in estimating the effect of viewing live-streamed content on game playing. Ideally, we would randomly assign users to treatment and control groups, restrict the control group's access to live-streamed content, and measure the difference in subsequent gameplay between the two groups. However, such an experiment is infeasible for commercial and ethical reasons. The core empirical challenge is that viewing and playing may be correlated with unobservable factors, making causal inference difficult. For instance, a spike in interest in a particular video game at a specific moment may lead to both increased gameplay and viewing of that game's live-streamed content.

To identify our treatment effect, we use an unexpected Twitch.tv interruption that lasted approximately 10 hours from August 30 to 31, 2017. This disruption is useful because it creates exogenous variation in users' ability to view live-streamed content without affecting their ability to play the game. Critically, while the event was system-wide, the server outage occurred at different times of day across the globe. The outage occurred during the afternoon and evening hours in the U.S. (Eastern Time), corresponding to late evening or past midnight in Europe. This geographic variation is crucial because users are generally more likely to watch live-streamed content and play video games during afternoons and evenings. Therefore, the event caused "greater" disruptions in livestreaming viewership for users in the U.S. compared to those in Europe.

The Twitch incident began around 3:00 PM on Wednesday, August 30, 2017 (U.S. Eastern Time), when users started experiencing issues accessing the website. Twitch released its first announcement via its support account on Twitter at 3:54 PM, acknowledging the technical issue affecting page loading and stating that its team was actively working on the problem. Ultimately, the technical issue persisted for about ten hours and was resolved around 1:00 AM on Thursday, August 31, 2017. In addition to several tweets made by the Twitch's support account, there was a significant spike in Google Trends for "Twitch Down," and media outlets reported that Twitch was unavailable from August 30 to 31, 2017 (Tylwalk 2017). We provide exhibits documenting the incident in Web Appendix B.

The event timeline across different countries is provided in Table 3. The exhibit reports the timing of the Twitch server outage across the different time zones where we sampled users and highlights (in dark grey) whether the downtime overlapped with usual entertainment hours, defined as 4 PM to midnight. As shown in Web Appendix C, gaming activity is most pronounced during this period, with 55% of total usage occurring within these eight hours. Most regions in the U.S. (except Hawaii and Alaska) experienced the outage during the afternoon and evening, whereas most European countries experienced it primarily in the late evening or past midnight.

Table 3: Service Status of Twitch and Local Times in Countries on August 30 and 31, 2017

| Time zone | US Pacific | US Mountain | US Central | US Eastern | EU Western | EU Central | EU Eastern |
|---|---|---|---|---|---|---|---|
| UTC time | UTC – 7h | UTC – 6h | UTC – 5h | UTC – 4h | UTC +1h | UTC +2h | UTC +3h |
| Countries | United States | | | | United Kingdom | France; Germany; Spain; Poland | Romania; Türkiye; Ukraine |
| Service status | | | | | | | |
| Normal | 9:00 AM | 10:00 AM | 11:00 AM | 12:00 PM | 5:00 PM | 6:00 PM | 7:00 PM |
| Normal | 10:00 AM | 11:00 AM | 12:00 PM | 1:00 PM | 6:00 PM | 7:00 PM | 8:00 PM |
| Normal | 11:00 AM | 12:00 PM | 1:00 PM | 2:00 PM | 7:00 PM | 8:00 PM | 9:00 PM |
| Down | 12:00 PM | 1:00 PM | 2:00 PM | 3:00 PM | 8:00 PM | 9:00 PM | 10:00 PM |
| Down | 1:00 PM | 2:00 PM | 3:00 PM | 4:00 PM | 9:00 PM | 10:00 PM | 11:00 PM |
| Down | 2:00 PM | 3:00 PM | 4:00 PM | 5:00 PM | 10:00 PM | 11:00 PM | 0:00 AM |
| Down | 3:00 PM | 4:00 PM | 5:00 PM | 6:00 PM | 11:00 PM | 0:00 AM | 1:00 AM |
| Down | 4:00 PM | 5:00 PM | 6:00 PM | 7:00 PM | 0:00 AM | 1:00 AM | 2:00 AM |
| Down | 5:00 PM | 6:00 PM | 7:00 PM | 8:00 PM | 1:00 AM | 2:00 AM | 3:00 AM |
| Down | 6:00 PM | 7:00 PM | 8:00 PM | 9:00 PM | 2:00 AM | 3:00 AM | 4:00 AM |
| Down | 7:00 PM | 8:00 PM | 9:00 PM | 10:00 PM | 3:00 AM | 4:00 AM | 5:00 AM |
| Down | 8:00 PM | 9:00 PM | 10:00 PM | 11:00 PM | 4:00 AM | 5:00 AM | 6:00 AM |
| Down | 9:00 PM | 10:00 PM | 11:00 PM | 0:00 AM | 5:00 AM | 6:00 AM | 7:00 AM |
| Normal | 10:00 PM | 11:00 PM | 0:00 AM | 1:00 AM | 6:00 AM | 7:00 AM | 8:00 AM |
| Normal | 11:00 PM | 0:00 AM | 1:00 AM | 2:00 AM | 7:00 AM | 8:00 AM | 9:00 AM |
| Normal | 0:00 AM | 1:00 AM | 2:00 AM | 3:00 AM | 8:00 AM | 9:00 PM | 10:00 PM |

Note: 0:00 AM in each column marks the beginning of August 31, 2017. Cells are color-highlighted: light grey cells denote times of streaming service interruptions, while dark grey cells indicate typical gaming hours (4 PM—11:59 PM) in specific time zones that overlapped with the down event. UTC times have been adjusted for daylight saving time during August 2017, where applicable.

## 4. Empirical Analyses

In this section, we explain how we identify the causal effect of live-streamed content on gameplay using the exogenous shock discussed in the previous section. We first estimate the direct effect of the Twitch server outage on users' gameplay minutes across different time zones. This provides initial empirical evidence that the shock disproportionately affected users' gameplay, providing us with an opportunity to use the server-down event to identify the effect of interest. Then, using a "server-down" dummy and the overlap of this shock with individual users' prime entertainment hours as instruments, we estimate the causal effect of livestreaming viewing minutes on gameplay minutes using a control function approach.

### 4.1. Panel Data and Summary Statistics

For our empirical analyses, we construct panel data for the five-week main analysis period. Since we have granular data on both live-streamed content viewing and gaming activities, and the exogenous shock lasted only about ten hours, we constructed a panel dataset at the user, day, and three-hour block levels. Specifically, we divided each calendar day into eight three-hour blocks. For each observation, we recorded whether the user played the video game and the duration of play. Given that the average duration of typical gameplay sessions is about 66 minutes (median 55.9 minutes), we refrain from scaling the panel data to the hourly level to avoid creating many censored observations (though we confirm robustness to this choice using one-hour panel in Web Appendix D). Summary statistics are reported in Table 4.

Table 4: Descriptive Statistics—Main Analysis Period

| Variables | N | Mean | Std. | $10^{th}$ | $50^{th}$ | $90^{th}$ |
|---|---|---|---|---|---|---|
| Gameplay (binary) | 2,837,240 | .162 | .368 | .000 | .000 | 1.000 |
| Play minutes | 2,837,240 | 10.690 | 30.820 | .000 | .000 | 41.350 |
| Play minutes \| Gameplay=1 | 458,687 | 66.121 | 47.012 | 14.117 | 55.900 | 137.600 |
| Stream view minutes | 2,837,240 | 6.522 | 24.633 | .000 | .000 | 9.450 |
| Stream view minutes \| Stream view minutes >0 | 361,049 | 51.253 | 49.755 | 5.000 | 32.780 | 132.440 |
| User tenure (logged weeks) | 2,837,240 | 4.674 | .527 | 3.949 | 4.816 | 5.191 |

4.2. Geography-based Analysis

Our first analysis examines whether users' gameplay minutes changed in response to Twitch's interruption. Suppose there is a statistically meaningful complementary relationship between live-streamed content viewing and gameplay. In that case, the server-down event should result in reduced gameplay, particularly in regions where the outage overlaps with peak entertainment hours. Conversely, if playing and viewing are substitutes, then we would observe an increase in gameplay.

The goal of the analysis is to generate empirical evidence for the use of the exogenous shock at the geographic level (see Seiler et al. 2017 for a similar analytical approach). It establishes the rationale for the key identification source used in subsequent analyses: variation in users' exposure to the Twitch outage creates exogenous differences in access to live-streamed content that are correlated with gameplay. To accomplish this, we estimate the following fixed effects regression:

$$\ln(\text{Playmins}_{ith} + 1) = \beta_1 Down_{c(i)th} + \beta_2 Down_{c(i)th} \times US_{c(i)} + \beta_3 Tenure_{it} + \alpha_i + \tau_w + \eta_{dh} + \varepsilon_{ith} \quad (1)$$

where $\text{Playmins}_{ith}$ denotes the play minutes recorded for user $i$ at the $h$-th three-hour block on day $t$. $Down_{c(i)th}$ is a binary indicator representing whether Twitch was unavailable (1) or not (0) in user $i$'s country $c$. The three-hour block takes a value of one if at least two hours within that block were affected by the server interruption. We also confirm the robustness of our findings by defining the block as affected if any portion of the three-hour block was impacted by the interruption (Web Appendix D). Since we do not have detailed time zone information within each country, we treated all users in the U.S. as being in the Eastern Time Zone (UTC-4), as it is the most populated time zone in the country, covering approximately 48% of the population (RPS Relocation 2018). To ensure robustness, we conducted a sensitivity analysis (see Web Appendix D) by randomly assuming that some users within the U.S. were in the Pacific Time Zone (UTC-7). $US_{c(i)}$ is a binary indicator that denotes whether the user is from the United States. The interaction between the server-down indicator and the U.S. country dummy captures any disproportionate impact of the server-down event. $Tenure_{it}$ represents the log of weeks since sign-up. We also include user-level fixed effects ($\alpha_i$), weekly time fixed effects ($\tau_w$), and day-of-week by three-hour block fixed effects ($\eta_{dh}$). $\varepsilon_{ith}$ is the error term.

Table 5: Impact of Twitch Service Interruption on Gameplay Minutes

| Model | (1) | (2) | (3) | (4) | (5) |
|---|---|---|---|---|---|
| Source of identification | Simple Difference | US vs. Non-US | Time from UTC hour | Google search volume of 'Twitch Down' | Overlapping hours of prime game hours |
| Moderator variable | | $\mathrm{US}_{c(i)}$ | $\mathrm{UTC\ Time}_{c(i)}$ | $\mathrm{SearchVolume}_{c(i)}$ | $\mathrm{AffectedHours}_{c(i)}$ |
| $\mathrm{Down}_{ith}$ | -.110$^{***}$ | -.052$^{***}$ | -.112$^{***}$ | -.112$^{***}$ | -.112$^{***}$ |
| | (.010) | (.015) | (.010) | (.010) | (.010) |
| $\mathrm{Down}_{ith}$ × Moderator | | -.099$^{***}$ | .016$^{***}$ | -.003$^{***}$ | -.019$^{***}$ |
| | | (.020) | (.003) | (.001) | (.004) |
| $\mathrm{Tenure}_{it}$ | -.439$^{**}$ | -.440$^{**}$ | -.440$^{**}$ | -.440$^{**}$ | -.440$^{**}$ |
| | (.145) | (.145) | (.145) | (.145) | (.145) |
| Point estimate for users in the U.S. | | -.150$^{***}$ | -.149$^{***}$ | -.139$^{***}$ | -.133$^{***}$ |
| | | (.014) | (.014) | (.013) | (.012) |
| R squared | .165 | .165 | .165 | .165 | .165 |
| R squared (adj.) | .162 | .162 | .162 | .162 | .162 |

Note: *p<.05; **p<.01; ***p<.001. The sample consists of 2,837,240 observations, with standard errors clustered at the player level (in parentheses). All models include user, week, and day-of-week × three-hour block fixed effects. UTC time, Google search volume, and affected hours are centered on their means.

We report the estimation results in Table 5. Column 1 shows a simplified model that only includes the server-down dummy. This simple difference model captures the average effect of the service incident across all users compared to normal operation days. We find that the server-down event significantly reduces gameplay minutes ($\beta$ = -.110; p<.001), indicating that even though the video game continued to operate normally during the event, users played the game less than usual. This provides preliminary evidence that live-streamed content and gameplay are complementary products.

In column 2, we estimate the model described in Equation (1) to determine whether the average effect identified in column 1 is more substantial in the United States. We find a negative and significant interaction effect ($\beta$ = -.099; p < .001), indicating that the impact of the Twitch disruption is more pronounced in the United States. The main effect of the server-down dummy remains negative and significant ($\beta$ = -.052; p < .001), suggesting that users from Europe also reduced their gameplay minutes in response to the Twitch interruption. This disproportionate impact aligns with our intuition that the event overlapped with more popular entertainment hours in the U.S. According to the calibration period data, about 64% of gameplay in the U.S. (Eastern time) occurs during the hours affected by the Twitch interruption. In contrast, a smaller proportion of gameplay takes place during the same hours in European countries, such as 37% in Spain, 40% in France, and 32% in Ukraine.

In columns 3 through 5, we report analyses using different specifications to capture the disproportionate effects of the interruption. First, we assign each country a UTC offset value based on its time zone (column 3). For example, during the observation period, the U.K.'s time zone is UTC+1, so we assign a value of one to users from this country. For the United States, we assume all users are in the Eastern Time Zone (UTC-4), assigning a value of -4. As the value increases, it indicates locations further east. Second, to account for the possibility that the interruption had a disproportionate impact, we use country-level Google Trends scores for the keyword "Twitch down" during August 30-31, 2017, and assign the country-level search score as a moderator (column 4). The rationale is that higher search intensity suggests the outage was more impactful. Third, we calculate the overlap between the prime entertainment hours (4 PM to midnight) and the Twitch interruption at the country level, assigning the number of overlapped hours accordingly (column 5). This is represented by the number of dark cells highlighted in Table 3 by country.

Across all three models, we consistently find a larger reduction in gameplay among U.S. users compared to European users. Furthermore, the positive interaction effect in column 3 indicates that the negative impact of the Twitch disruption diminishes as the region is farther east of the U.S. Table 5 includes a row labeled "Point estimate for users in the U.S.," which summarizes the adjusted effect for U.S. users based on the interaction terms. Across columns 2 to 5, we find consistent and similar effect sizes of the Twitch disruption on American users, showing approximately a 15% reduction in gameplay minutes. The results in Table 5 consistently demonstrate that when live-streamed content becomes inaccessible, users tend to *reduce* their gameplay minutes.

### 4.3. Effect of Live-Streamed Content Viewing on Gameplay

Next, we investigate the impact of viewing live-streamed content on gameplay. Estimating the causal effect of livestreaming on gaming requires careful consideration, as both behaviors may be caused by the same unobserved factors. In particular, unobservable shocks to a user's available free time can simultaneously influence livestreaming and gaming behaviors, a concern commonly referred to as activity bias (Lewis et al. 2011).

To address this challenge, we use the unexpected global Twitch outage and users' geolocation (i.e., country information) as sources of exogenous variation. Identification is based on the relationship between users' gameplay minutes and their differential exposure to Twitch service disruptions during their prime entertainment hours across countries.

The preceding analyses accounted for heterogeneous effects across countries. We next consider how individual differences may also create heterogeneity in the disruption's impact. To account for within-country heterogeneity in disruption exposure, we also use individual differences in the overlap between each user's prime entertainment hours and the local timing of the outage. This step allows us to utilize variation not only across countries but also across individuals within the same country who experienced different levels of exposure to the disruption due to differences in their preferred entertainment hours.

To accomplish this, we identify individual users' top entertainment hours ($\text{AffectedHours}_i$) from their gaming logs during the four-week calibration period. We count the number of matches played in each hour, and we identify the top eight hours with the highest gameplay activity for each user, treating these as their most preferred entertainment hours. We then calculate the number of hours that overlap with the local time of the Twitch disruption. By construction, the overlap hours can range from zero to eight. The identifying variation in this overlap stems from the coincidence between an outage whose timing is externally fixed and a user's habitual active hours: two users with identical habitual hours can receive very different values depending solely on their local time zone. Identification therefore comes from this exogenous coincidence, driven by geography and event timing, rather than from a user's time preference itself. Our underlying assumption is that the overlap between Twitch downtime and the individual's habitual active hours varies across countries, but generally increases as the region gets closer to the U.S. and decreases as it gets farther away. Figure 3 illustrates the distribution of affected hours across different countries.

Figure 3: Overlapping Hours with Twitch Down Among Top Eight Preferred Entertainment Hours

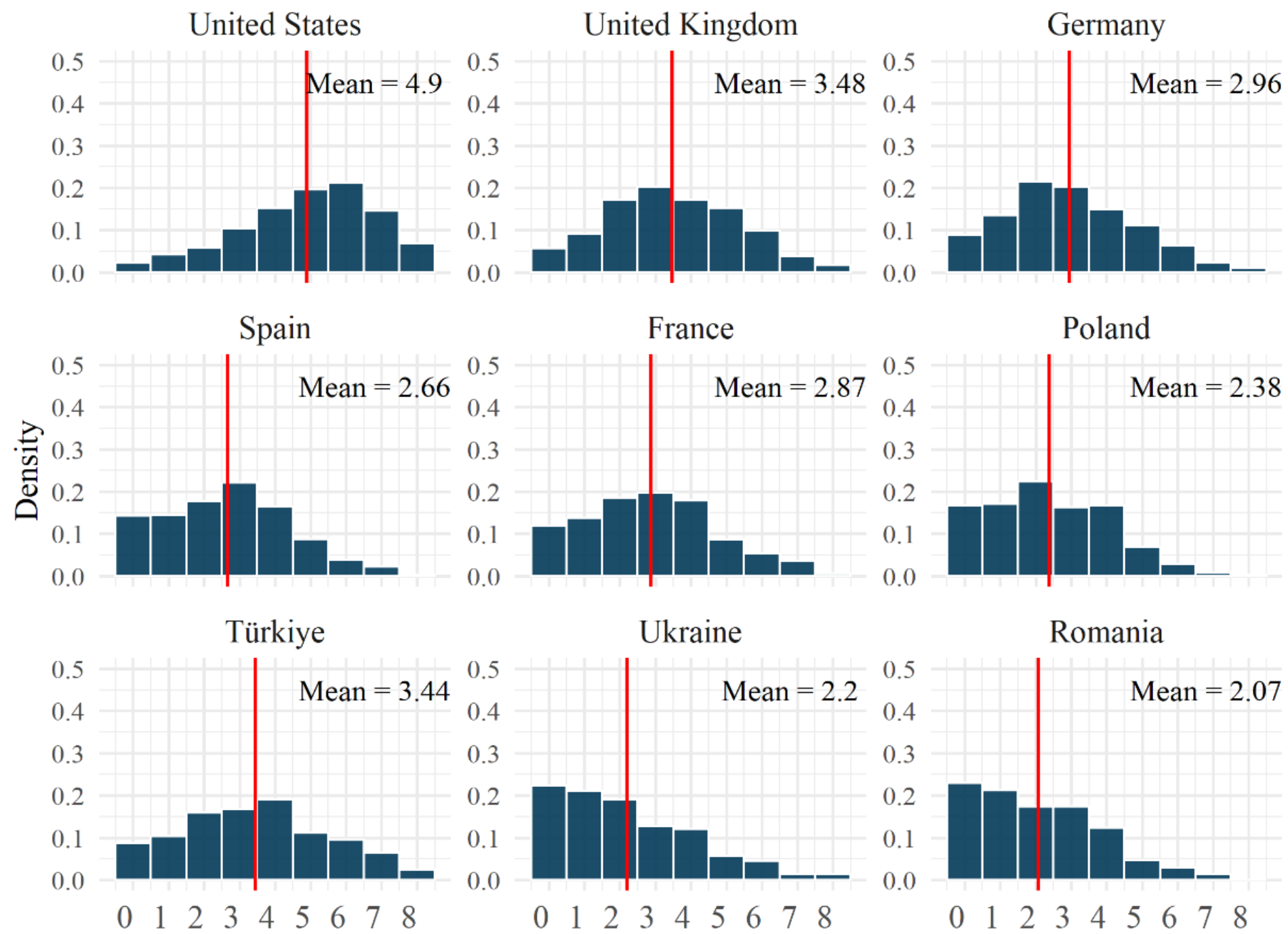

Note: Color is available online. Red vertical bars indicate the positions of country-level means.

The patterns in Figure 3 confirm our expectation that U.S. users were more affected by the Twitch disruption than those in Europe. The calculations also provide individual-level information on how much of their preferred entertainment hours were affected by the Twitch service outage. This information may improve identification by allowing us to capture individual differences within each country, based on time-preference-related shocks from the Twitch interruption. We also observe that the distribution of overlapping hours becomes increasingly right-skewed as the country's time zone is farther from the U.S. This indicates that users in these countries were less affected, as the Twitch downtime did not significantly overlap with their most preferred gameplay hours.

To estimate the causal effect using this identification strategy, we employ a control function approach (Wooldridge 2015, Papies et al. 2017). The control function approach is similar to traditional instrumental variable (IV) regression in that both utilize exogenous variation from instrumental variables. However, it provides greater flexibility, particularly when interactions involve potentially endogenous variables. This is crucial for our analysis, as we wish to understand how the effect of live-streamed content on gameplay interacts with users' category viewership shares across different types of streamers. The control function approach involves obtaining the residuals from a first-stage regression and then including these residuals as an additional regressor in the second-stage regression. This included residual, or control function, not only addresses the bias of the endogenous variable (Petrin and Train 2010) but also corrects for potential endogeneity issues for interaction terms (Wooldridge 2015).

We specify the first-stage regression using a framework similar to the previous section. To predict the endogenous variable, live-streamed content viewing minutes, we use two instruments: (1) a dummy variable indicating when the Twitch outage occurred at the user's local time ($Down_{c(i)th}$), and (2) an interaction between this outage dummy and the number of the individual user's preferred entertainment hours affected by the outage ($Down_{c(i)th} \times \text{AffectedHours}_i$). The variable $\text{AffectedHours}_i$ is centered on its mean to facilitate the interpretation of the main effect at the average level of overlap. Equation (2) describes the first-stage model. We obtain the residuals from the first-stage model and include them as an additional regressor ($\widehat{\omega}_{ith}$) in the model where we estimate the effect of live-streamed content viewing on gameplay minutes (Equation 3). The full estimation system is given below.

***First-stage model:***

$$\ln(\text{Viewmins}_{ith}+1) = \gamma_1 Down_{c(i)th} + \gamma_2 Down_{c(i)th} \times \text{AffectedHours}_i + \gamma_3 Tenure_{it} + \alpha_i + \tau_w + \eta_{dh} + \omega_{ith} \tag{2}$$

***Second-stage model:***

$$\ln(\text{Playmins}_{ith}+1) = \delta_1 \ln(\text{Viewmins}_{ith}+1) + \delta_2 Tenure_{it} + \widehat{\omega}_{ith} + \alpha_i + \tau_w + \eta_{dh} + \varepsilon_{ith} \tag{3}$$

Most variables remain the same as in Equation (1). $\text{Viewmins}_{ith}$ represents the view minutes of live-streamed content recorded for user $i$ during the $h$-th three-hour block on day $t$. Theoretically, $\gamma_1$ captures the average impact of the Twitch interruption on all users when $\text{AffectedHours}_i$ is at its mean, while $\gamma_2$ captures how the effect varies with the number of preferred entertainment hours that overlap with the outage. Finally, $\widehat{\omega}_{ith}$ denotes the residuals from the first-stage regression (i.e., control function), which adjusts for the potential endogeneity of the livestreaming viewing variable.

Potential IVs need to satisfy five conditions to identify the effect of an endogenous regressor (Grewal and Orhun 2024): independence, stable unit treatment value assumption (SUTVA), relevance, monotonicity, and exclusion restriction. First, independence is plausible because the Twitch outage was unexpected and not triggered by user-specific behaviors. Second, SUTVA, which requires that one unit's treatment or IV does not affect another unit's potential outcomes, is also reasonably satisfied. In our context, this means that a user's gameplay behavior is not influenced by changes in other users' exposure to the IV. This holds because individual users' viewing minutes are not publicly revealed on the platform, and the game does not display real-time player statistics, minimizing the likelihood of spillovers or externalities. In addition, because the game relies on a large, global matchmaking pool, variation in users' exposure to the Twitch outage across time zones does not meaningfully affect other users' ability to find matches or play the game, further supporting the non-interference assumption. Third, relevance is straightforward: the outage caused measurable variation in live-streamed content viewing, and this impact increases for users whose preferred entertainment hours overlapped more with the outage window. Fourth, this directional pattern also supports the monotonicity condition (i.e., the IV affects the treatment in the same direction for all units), as the outage consistently reduces streaming across users, with larger reductions for those whose prime entertainment hours overlap more with the outage window. We show below in empirical results that we find the expected negative and significant effects of the outage and its interaction term, consistent with the instruments' relevance. Fifth, the exclusion restriction is likely to hold, as the outage affected access to live-streamed content but did not disrupt gameplay, thus influencing gameplay only through changes in exposure to streaming.

A potential concern for the exclusion restriction is that the outage disabled all of Twitch, so users who normally watch non-focal content (e.g., other games or Just Chatting) might redirect that freed time directly into the focal game. If this is the case, however, the outage would *increase* gameplay, whereas we observe a *decrease* (Table 5). This scenario therefore works against, rather than produces, our estimated complementarity (main effect), leaving our estimates conservative.

Table 6: Elasticity of Gameplay Minutes Relative to Live-Streamed Content View Minutes

| Model | First-stage | | Second-stage | |
|---|---|---|---|---|
| Dependent variable | $\ln(\text{Viewmins}_{ith}+1)$ | | $\ln(\text{Playmins}_{ith}+1)$ | |
| Variables | Est. | S.E. | Est. | S.E. |
| $\ln(\text{Viewmins}_{ith}+1)$ | | | .343*** | .036 |
| $\text{Down}_{c(i)th}$ | -.264*** | .007 | | |
| $\text{Down}_{c(i)th} \times \text{AffectedHours}_i$ | -.015*** | .003 | | |
| $\text{Tenure}_{it}$ | -.332** | .112 | -.325* | .143 |
| $\widehat{\omega}_{ith}$ (control function) | | | -.160*** | .036 |
| F-statistic of: | | | | |
| $\text{Down}_{c(i)th}$ | 1,265.18*** | | | |
| $\text{Down}_{c(i)th} \times \text{AffectedHours}_i$ | 23.01*** | | | |
| R squared | .192 | | .183 | |
| R squared (adj.) | .189 | | .180 | |

Note: *p<.05; **p<.01; ***p<.001. The sample consists of 2,837,240 observations. All models include user, week, and day-of-week × three-hour block fixed effects. First-stage standard errors are clustered at the user level. Second-stage bootstrapped standard errors are based on 50 resamples. The affected hours variable is mean-centered.

Table 6 reports the estimation results of our two-stage model. The first-stage model results show that the unexpected interruption resulted in a significant decrease in live-streamed content viewing time. The effect was more pronounced among users whose preferred entertainment hours overlapped with the Twitch unavailability period. This implies that the impact of the Twitch downtime is not uniform across all users but is amplified for those whose prime entertainment hours are more severely affected by the interruption. This suggests that gaming and viewing behaviors are closely interlinked. We find that our instruments are empirically valid, as the F-statistics for both variables are statistically significant and exceed the commonly recommended threshold of 10 for assessing instrument strength.

In the second-stage model, we confirm that viewing live-streamed content has a positive and statistically significant effect on play minutes. The estimated elasticity is .343, which implies that a 10% increase in live-streamed content viewing minutes results in approximately a 3.43% increase in gameplay minutes. We also find that the estimate for the control function is statistically significant, indicating that endogeneity bias is present and is controlled for. The negative estimate of the control function suggests that unobserved factors that increase viewing time are associated with fewer play minutes.

### 4.4. Processes: How Exactly Does Livestreaming Complement Gameplays?

The previous analysis confirms the positive effect of livestreaming on gameplay. We next investigate *how* live-streamed content influences the play of featured games and consider two possible processes by which livestreaming can complement gameplay. One way live-streamed content can increase gameplay is by encouraging people to enjoy both activities ***concurrently***. Concurrent consumption means that individuals may play the game while simultaneously watching someone else's gameplay on the livestreaming platform. Anecdotal evidence suggests that some users prefer concurrent consumption (Ashworth 2022), a practice similar to using smartphones while watching television (commonly referred to as "second screening") (Van Cauwenberge et al. 2014). To examine this possibility, we merge the viewing and gameplay logs and calculate the ratio of minutes users spent on both activities to their total minutes spent viewing live-streamed content. A ratio of zero indicates no concurrent consumption, while a ratio of one indicates complete concurrent consumption. We find that users' concurrent consumption rates have many nonzero values but are quite low overall. The median rate is 4.28%, with an interquartile range from 0% (25th percentile) to 13.78% (75th percentile).

Alternatively, streamed content may affect gameplay through users playing the game after viewing. For instance, some users may have routines where they watch live-streamed content first, finish their viewing, and then play the game afterward. If this is the case, live-streamed content leads to sequential rather than simultaneous video game play. To examine this possibility, we track each user's viewing sessions during the calibration period. We examine these consumption patterns during the calibration period, which is free of the outage, so that the observed routines reflect normal behavior rather than shock-induced disruption. On average, users spend 38 minutes in each viewing session. From the end timestamp of each viewing session, we recorded whether there were any follow-up video game plays within three hours (180 minutes).

Figure 4: Follow-Up Gameplay Rates Within a Three-Hour Window

| ***Panel A:*** *Gameplay Rates: Viewing vs. Non-Viewing* | ***Panel B:*** *Cumulative Follow-Up Gameplay Rates* |
| --- | --- |

Figure 4 reports the results of our descriptive analysis. Panel A compares gameplay rates (%) within a three-hour window, with and without livestreaming viewing. For the baseline condition, we compute gameplay rates for three-hour panel blocks on days with no viewing activity. For the after-viewing condition, we calculate gameplay rates following all viewing sessions within a three-hour window. The difference is substantial: users play the game after viewing live-streamed content (30.9%) at almost twice the rate compared to days without viewing activity (15.1%). Panel B shows more detail of sequential gameplay within three hours of viewing. Gameplay rates increase sharply, reaching over 20% within the first hour after viewing streamers. This already exceeds the baseline gameplay rate on non-viewing days, providing clear evidence of follow-up gameplay. After the first hour, the rate continues to rise, eventually surpassing 30% by the three-hour mark. Collectively, the two panels of Figure 4 provide evidence of substantial sequential consumption.

The preceding analysis shows that livestreaming can complement gameplay through both concurrent and sequential processes. However, our findings suggest that sequential consumption is the more dominant form. In fact, the majority of follow-up gameplay after viewing occurs within the first hour. This provides empirical evidence that livestreaming primarily complements broadcast video game content in a sequential manner, motivating timely follow-up gameplay.

### 4.5. Streamer Type Viewership and User-level Response Heterogeneity

Our analyses have addressed several questions: whether, to what extent, and how exactly live-streamed content supports the playing of a featured video game. The results suggest that investments in partnerships with creators on livestreaming platforms enhance a game's engagement – there is overall more playing and viewing triggers subsequent playing. Another question is how these positive effects vary based on the "types" of streamers with whom a player engages. A micro streamer focused on a specific game may play a different role within a game's fandom community than a mega streamer with significant production values who monetizes his own audience. Differences in streamers' audience structures and monetization goals highlight managerially important questions about streamer heterogeneity and influence. Specifically, we are interested in how the type of streamer a user watches moderates its influence.

Popularity, as measured by audience size, may affect influence through multiple mechanisms. It may signal the performer's potential for influence (Lu et al. 2021) as larger audiences provide greater exposure to a sponsor (Gu et al. 2024). At the consumer level, mega streamers attracting thousands of viewers might be perceived differently in terms of source credibility than micro streamers with audiences in the dozens (Sternthal et al. 1978). Differences in audience sizes between mega and micro streamers may also affect the nature and frequency of performer-viewer interactions (Gu et al. 2024). As audience size grows, streamers are less able to interact with audience members (Ford et al. 2017, Lu et al. 2021, Gu et al. 2024). Limited interaction with mega streamers may weaken the mega streamer-viewer relationship and, consequently, reduce influence (Fournier 1998). In addition to the obvious structural challenge of interacting with individuals in an audience of thousands rather than tens, researchers have also found that audience size increases consumers' monetary costs of interaction (Lu et al. 2021). In contrast, micro streamers' smaller audiences may offer a more relationship-focused viewing experience, more akin to enjoying the game with close friends.

In addition to audience size, other streamer attributes may affect the performer's ability to influence viewers. In particular, the relationship of the streamer channel to the promoted game may affect influence. The publisher's channel may be an especially credible source because it has access to inside information about game updates or official esports competitions. Official channels may also appear more professional, as they often have larger budgets and higher production values. However, if viewers see the channel as a marketing tool, they may view it skeptically.

We speculate that different fundamentals may drive viewing concentration patterns across streamer types. Mega streamers may offer celebrity appeal and broadly shared experiences, while the publisher's channel may provide insider information and curated content. Micro streamers may provide niche content and high interactivity. The distinct benefits of each streamer type may have different downstream effects on subsequent player activities. We do not make directional predictions for shares a priori; instead, we allow the data to reveal the empirical patterns. Our strategy is to conduct post hoc analyses to better understand the mechanisms behind any observed heterogeneity.

While count-based metrics such as meaningful ties capture whether a user repeatedly engages with specific streamers, share-based metrics capture how a user's viewing time is allocated across streamer categories. We focus on viewership shares, as they directly capture the relative allocation of attention across streamer types. For our model operationalization, we combine repeat viewing with share of viewing. For each user–streamer pair that qualifies as a meaningful tie (Section 2.2), we calculate the share of viewing minutes relative to the user's total viewing minutes. We then aggregate these shares across meaningful streamers within the same category (firm-owned, mega, and micro). This procedure yields three category viewership share variables, corresponding to firm-owned, mega, and micro streamers. The remaining share of viewing time, allocated to streamers that do not meet the threshold for meaningful ties, is classified as non-relationship viewing. These category viewership share variables are constructed at the individual level (subscript $i$) based on viewing behaviors in the four-week calibration period immediately preceding the analysis period.

Table 7 presents summary statistics for our category viewership share measures across the three streamer types based on users who have at least one meaningful relationship with a streamer during the calibration period (Table 2). On average, users devote the largest fraction of their total viewing time to mega streamers (mean = .471), suggesting that repeated and meaningful viewing activities tend to concentrate in this category. Viewership shares for the publisher channel (mean = .082) and micro streamers (mean = .100) are lower and more dispersed than those of mega streamers, although the upper tails show a non-trivial subset of users who follow these types. To clarify the interpretation of Table 7, consider a user with at least one meaningful tie. On average, such a user allocates 8.2% of total viewing time to firm-owned channels, 47.1% to mega streamers, and 10.0% to micro streamers, with the remaining 34.7% directed to streamers without meaningful ties.

Table 7: Descriptive Statistics of Category Viewership Shares

| *Variables* | *Mean* | *Std.* | *5th* | *25th* | *50th* | *75th* | *95th* | *98th* | *99th* | *Max* |
|---|---|---|---|---|---|---|---|---|---|---|
| $ViewShare_i^{FirmOwned}$ | .082 | .201 | .000 | .000 | .000 | .000 | .559 | .903 | 1.000 | 1.000 |
| $ViewShare_i^{Mega}$ | .471 | .303 | .000 | .228 | .501 | .715 | .942 | .986 | .998 | 1.000 |
| $ViewShare_i^{Micro}$ | .100 | .228 | .000 | .000 | .000 | .031 | .711 | .935 | .983 | 1.000 |

Note: Sample includes 5,164 users with at least one meaningful relationship with streamers. Viewing behaviors are based on the four-week calibration period.

To examine heterogeneity in streamer type effects, we analyze how the impact of live-streamed content on gameplay varies across users based on the types of streamers they repeat watch. Specifically, we relate this variation to users' category-level viewership shares across firm-owned, mega, and micro streamers. These variables were mean-centered and interacted with live-streamed content viewing minutes. Equation (4) extends the second-stage specification in Equation (3) by interacting live-streamed content viewing minutes with the mean-centered category viewership shares, while retaining the same control function correction ($\widehat{\omega}_{ith}$) obtained from the first stage in Equation (2).

$$\begin{aligned}\ln(\text{Playmins}_{ith}+1) = \; & \delta_1 \ln(\text{Viewmins}_{ith}+1) \\ & + \delta_2 \ln(\text{Viewmins}_{ith}+1) \times ViewShare_i^{FirmOwned} \\ & + \delta_3 \ln(\text{Viewmins}_{ith}+1) \times ViewShare_i^{Mega} \\ & + \delta_4 \ln(\text{Viewmins}_{ith}+1) \times ViewShare_i^{Micro} \\ & + \delta_5 Tenure_{it} + \widehat{\omega}_{ith} + \alpha_i + \tau_w + \eta_{dh} + \varepsilon_{ith}\end{aligned} \quad (4)$$

Table 8 reports the estimation results from Equation (4). To test the robustness of our definition of meaningful user-streamer ties (at least eight visits, column 2), we replicate the results using relaxed (4 visits, column 1) and tightened (12 visits, column 3) criteria. For simplicity, we focus on column (2), as the other columns yield consistent results. The results reveal several interesting patterns. First, we again find a positive and significant main effect of live-streamed content viewing on gameplay ($\delta_1$ = .340; p <.001). This effect represents the causal impact of viewing for users with average category-level viewership shares across the three streamer types.

Table 8: Heterogeneous Effects of Live-Streamed Content Viewing by Streamer Type Viewership Share

| Meaningful streamer-user tie is defined as: | ***4 repeated visits*** during the calibration period | ***8 repeated visits*** during the calibration period | ***12 repeated visits*** during the calibration period |
|---|---|---|---|
| Model | (1) | (2) | (3) |
| $\ln(\text{Viewmins}_{ith}+1)$ | .339*** | .340*** | .339*** |
| | (.036) | (.039) | (.036) |
| $\ln(\text{Viewmins}_{ith}+1) \times \boldsymbol{ViewShare_i^{FirmOwned}}$ | -.091*** | -.058*** | -.070* |
| | (.012) | (.017) | (.028) |
| $\ln(\text{Viewmins}_{ith}+1) \times \boldsymbol{ViewShare_i^{Mega}}$ | -.082*** | -.068*** | -.065*** |
| | (.011) | (.007) | (.008) |
| $\ln(\text{Viewmins}_{ith}+1) \times \boldsymbol{ViewShare_i^{Micro}}$ | .150*** | .176*** | .187*** |
| | (.019) | (.016) | (.026) |
| $\text{Tenure}_{it}$ | -.320* | -.326 | -.324* |
| | (.147) | (.167) | (.132) |
| $\widehat{\omega}_{ith}$ (control function) | -.155*** | -.155*** | -.155*** |
| | (.036) | (.040) | (.036) |
| R squared | .185 | .184 | .184 |
| R squared (adj.) | .182 | .181 | .181 |

Note: *p<.05; **p<.01; ***p<.001. The sample consists of 2,837,240 observations. All models include user, week, and day-of-week × three-hour block fixed effects. Bootstrapped standard errors with 50 resamples are used and they are in parentheses. The moderators are all centered on their means.

However, we find significant differences based on streamer type relationship preferences. The positive effect decreases when users devote a larger share of their viewing to the firm-owned channel ($\delta_2$

= -.058; p <.001) or mega streamers ($\delta_3$ = -.068; p <.001). In contrast, we find the opposite pattern among users whose viewership is more heavily concentrated on niche or small (micro) streamers ($\delta_4$ = .176; p <.001). The positive interaction indicates that live-streamed content is particularly effective in motivating gameplay when users' viewing shares are tilted toward micro streamers.

This pattern suggests that micro streamers, whose smaller audiences foster more interactive and community-oriented viewing environments, may be more effective at converting viewership into active game participation than mega streamers or the firm-owned channel that caters to larger, more anonymous audiences. This finding suggests a promising avenue for future research on how the community structures surrounding different streamer types shape downstream engagement with the products they feature.

To fully evaluate these adjusted effects within the range of observed data, we compute and plot the elasticity of live-streamed content viewing at varying levels of users' category-level viewership shares in Figure 5. Specifically, we use the results from model (2) in Table 8 to estimate the adjusted effect of live-streamed content on gameplay for three viewership-share profiles: (i) no meaningful relationship with this streamer category (when the original, non-centered category viewership share equals zero), (ii) low viewership share (25th percentile among positive category viewership shares), and (iii) high viewership share (75th percentile among positive category viewership shares).

Figure 5: Viewing–Gameplay Elasticity by Category Viewership Share

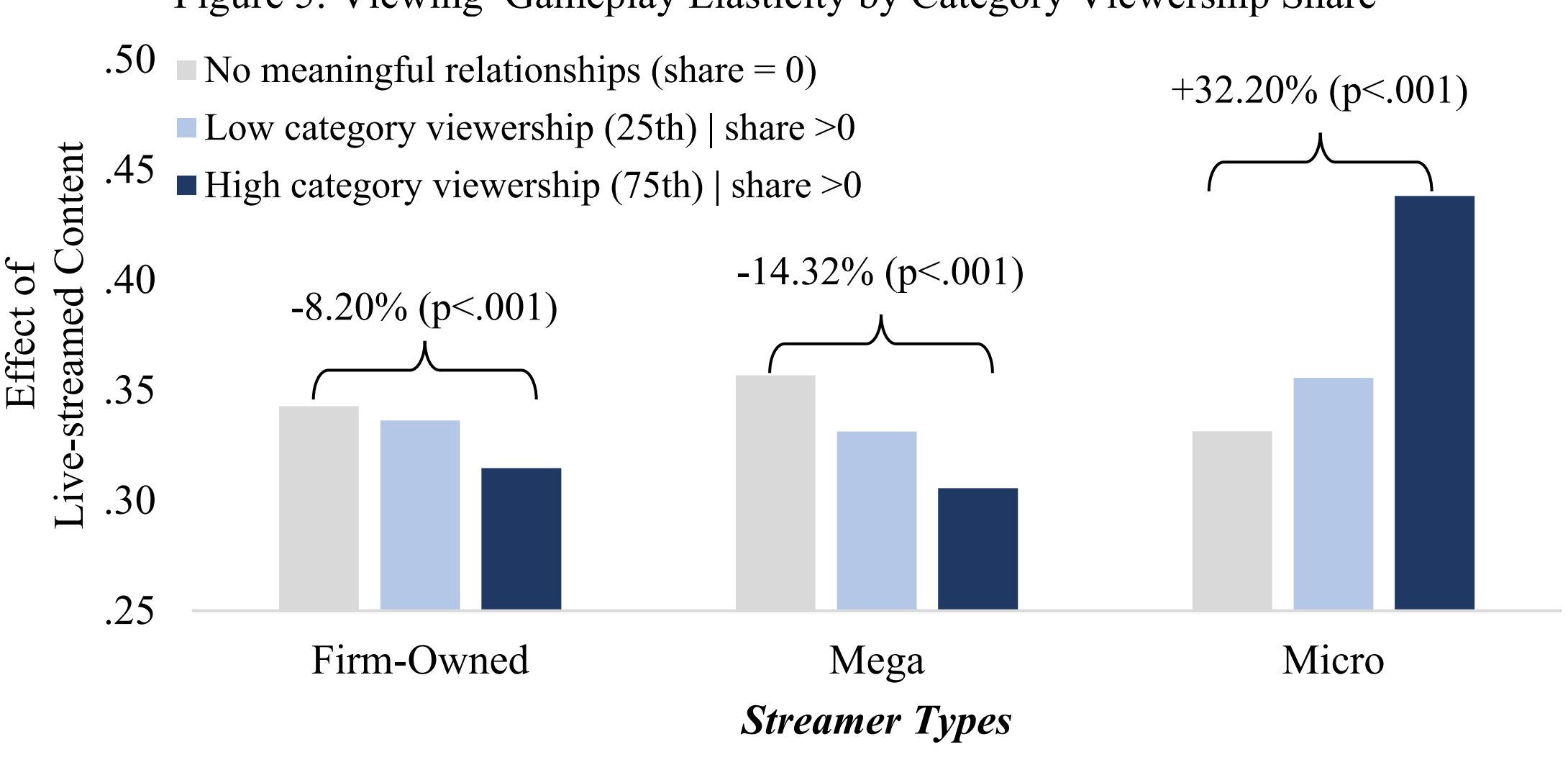

Note: Color is available online. The statistical significance of the adjustment (in %) between high and zero category-level viewership shares is shown.

The results show that the positive effect of live-streamed content on gameplay decreases as users' category-level viewership shares (viewers have more relationships with streamers with larger audiences) shift toward firm-owned channels and mega streamers, though the magnitude of decrease differs. Comparing no meaningful relationship to high category viewership share, the reduction is 8.20% for firm-owned channels and 14.32% for mega streamers. This highlights that repeat viewing of large independent creators is associated with a greater decrease in the positive spillover to the endorsed product.

However, the smaller reduction for the firm-owned channel suggests that this channel may still play a significant role in maintaining a connection between the brand and its audience. Since firm-owned channels can directly control the narrative and consistently deliver brand-focused content, they offer a strategic advantage in ensuring that viewership remains tied to the product, even with a slight reduction in effect size. This enables firms to maintain a direct and loyal customer base while fostering product engagement. The publisher's channel may even feature specialized content like esports that represents high production value, standalone content. This type of content may be more oriented to brand building and customer acquisition than short-term product promotion (Jo and Lewis 2024). The firm channel has incentives more aligned with community development. It has an interest in monetizing the community but this monetization is directly related to the game not an external human brand like a mega streamer.

In contrast, when users' viewing becomes more concentrated on mega streamers, the beneficial effects of watching live-streamed content are significantly less. Unlike micro streamers who remain embedded community members, mega streamers face stronger incentives to build their own personal brand, and their reduced capacity for interaction may further weaken the link to the game's community.

For micro streamers, we observe the opposite pattern: the positive effect increases as users' streamer relationships become more concentrated in this category. The adjusted effect increases by 32% when comparing the effect estimates between "no meaningful relationships" and "high category viewership". The contrasting pattern for micro streamers relative to firm and mega streamers is vital because it highlights how content created by streamers with niche and small to medium-sized fan bases can contribute positively to a fandom economy. Unlike firm-owned or mega streamers, micro streamers typically have small audiences, allowing them to produce focused or niche content while maintaining interaction with their viewers.

An important practical consideration is that our estimates reflect per-user elasticities, which do not account for differences in audience size across streamer types. While micro streamers drive stronger per-user responsiveness, mega streamers command substantially larger audiences, meaning their aggregate impact on total gameplay may still be greater. This implies that the optimal sponsorship strategy depends on the firm's objectives and costs of sponsorship: mega streamers may be more effective for maximizing broad exposure, whereas micro streamers may yield deeper engagement within a more targeted community. Balancing these goals and accounting for the relative costs of partnerships are key considerations for firms designing creator collaboration strategies.

### 4.6. Robustness Checks to Internal and External Validity

A potential issue with the generalizability of our findings is whether they are caused by something specific to the game. We believe this is unlikely because the game shares many characteristics with other games featured on livestreaming platforms, including real-time multiplayer mechanics, frequent content updates, and active engagement with the streaming community.

To ensure generalizability, we nonetheless replicate our analysis using an alternative methodology and a broader range of data. Specifically, we examine daily data on player counts and Twitch viewership for 268 other video game titles over the exact same observation period used in our main analysis (Aug 15 to September 18, 2017). These data were scraped with the website administrator's written permission from *SteamDB,* an independent platform that archives daily active game players and Twitch viewers. Consistent with our main analysis, for each game $g$ on day $t$, we observe the daily number of active players ($\ln(Players_{gt})$) and the daily number of Twitch viewers ($\ln(Viewers_{gt})$), both in logs. To isolate the causal effect of viewership on play, we employ the same control-function strategy used in our primary analysis, instrumenting $\ln(Viewers_{gt})$ with an indicator for the same Twitch outage interacted with the game's U.S. popularity. Our assumption is that games more popular with U.S. audiences are disproportionately exposed to the outage shock. The first-stage residual ($\hat{\varphi}_{gt}$) is then included as a control function in the second stage. The model includes individual game fixed effects ($\alpha_g$) and weekly time fixed effects ($\tau_w$), as specified in Equation (5).

$$\ln(Players_{gt}) = \eta_1 \ln(Viewers_{gt}) + \alpha_g + \tau_w + \hat{\varphi}_{gt} + \varepsilon_{gt} \quad (5)$$

As in our main analysis, we expect complementarity between livestreaming viewership and game participation and that the strength of the relationship varies with viewership concentration. Whereas the main analysis measured viewing share across three streamer types at the "individual" user level, the broader *SteamDB* sample offers many games but limited per-user granularity. We therefore shift to a game-level measure: $\%Top5_g$, the share of a game's Twitch viewership captured by its five most-watched channels in the month preceding the observation period, computed from *SullyGnome*, a third-party analytics platform that tracks streamer-level viewership statistics on Twitch. Importantly, $\%Top5_g$ is essentially uncorrelated with a game's overall popularity: its correlation with average daily active players over the observation period is −.016 (p = .79). This indicates that our viewership concentration measure captures the structure of a game's viewing community rather than its size.

We partition games into low-, medium-, and high-concentration terciles at the 33rd and 66th percentiles of $\%Top5_g$ and estimate the second-stage model separately for each tercile. This approach allows all coefficients, including the endogeneity correction, to vary freely across concentration groups without imposing functional form assumptions on the moderating relationship. For completeness, a pooled interaction specification with tercile indicators yields consistent results (Web Appendix E).

Figure 6 plots the estimated elasticity of daily active players with respect to daily Twitch viewers across the three tercile groups. We find consistent evidence of complementarity between livestreaming viewing and gameplay; however, the magnitude of this complementarity decreases as viewership becomes more concentrated among the top five mega streamers. This pattern indicates that the complementarity between livestreaming and gameplay weakens systematically as a game's viewing community becomes concentrated around a few mega streamers, and is strongest when viewership is less concentrated on these dominant channels. We also find that the median-split specification yields qualitatively similar results, with the top 50% group exhibiting significantly weaker complementarity than the bottom 50% group.

Figure 6: Complementarity between Livestreaming and Gameplay by Top Five Streamer Concentration

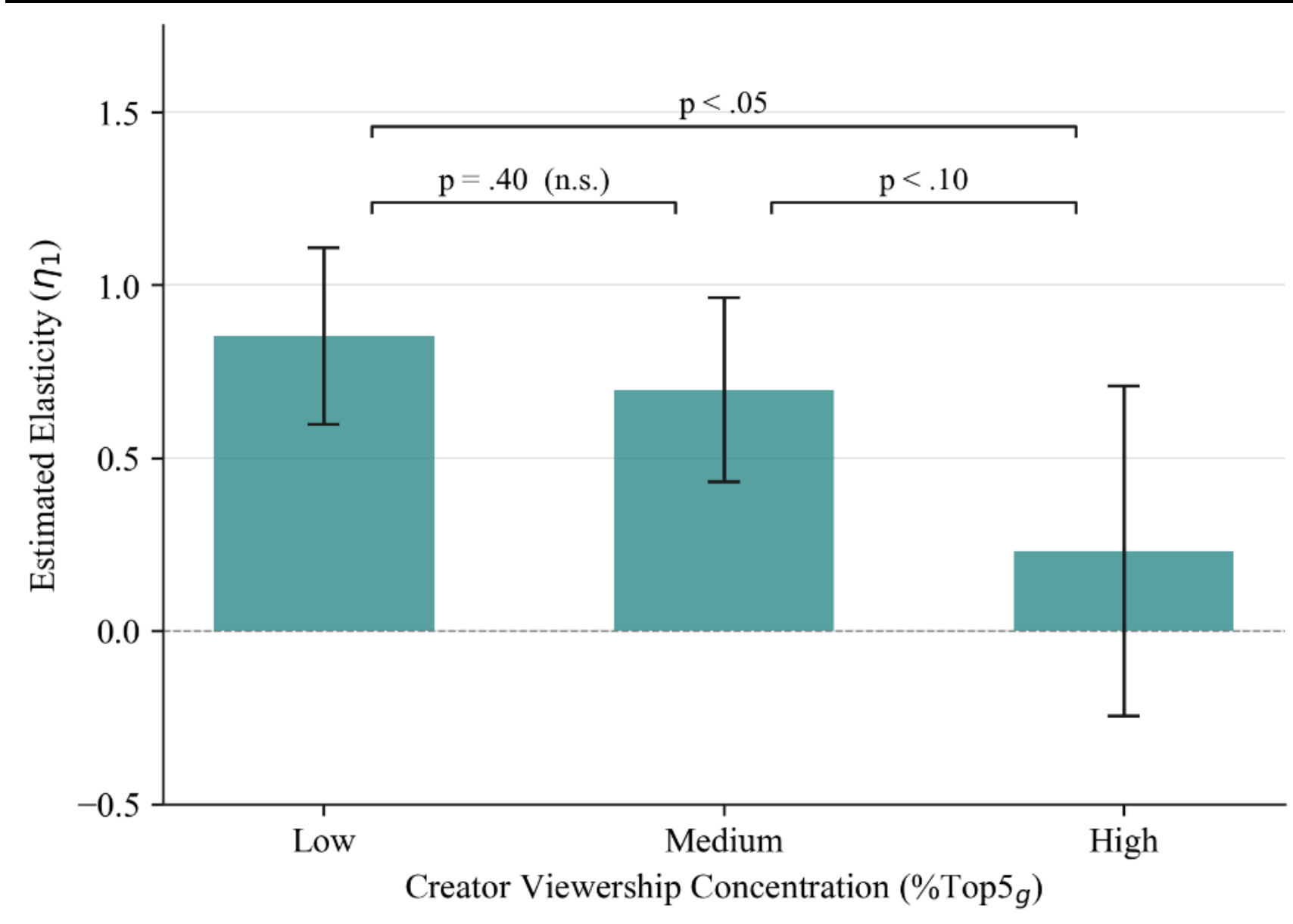

Note: Color is available online. Error bars denote 95% confidence intervals for the elasticity estimates.

This generalizability analysis complements our primary analyses but differs in important ways. While we employ a similar control function approach using the Twitch outage as an instrument, leveraging the assumption that games more popular among U.S. audiences were disproportionately affected, the aggregate nature of these data limits the strength of causal identification relative to our individual player-level analyses. Additionally, the daily game-level structure does not permit examination of individual co-consumption patterns or user-level heterogeneity by streamer type. Nonetheless, the consistent direction of results across 268 titles strengthens confidence in the generalizability of our core findings. Additional details on data construction and results are provided in Web Appendix E.

To assess the internal validity of our findings, we conducted a series of supplementary analyses. For instance, a reasonable concern regarding our empirical analyses is whether the results may be artifacts of specific operationalizations of certain variables. We replicated the analyses using alternative operationalizations of our key variables, including different assumptions about time zones in the U.S. sample and stricter or more lenient definitions of the viewership share measures. Because our dependent variable contains a high share of zero-valued observations, we further verify that the results are robust to a non-logged specification and to a Poisson fixed effects model that handles zeros more flexibly. The heterogeneity pattern across streamer types remains unchanged in both. For conciseness, we provide a summary of the supplementary analyses conducted in Table 9. A comprehensive list of all supplementary analyses, detailed methodologies, and results is available in the Web Appendix D.

Table 9: Robustness Analyses

| Robustness checks | Results available at: |
| --- | --- |
| • Alternative operationalization for “Down” dummy | WA-D.1 |
| • Alternative time zone assumption for U.S. sample | WA-D.2 |
| • Alternative operationalizations of affected hours | WA-D.3 |
| • Alternative operationalization for category viewership shares | Table 8 |
| • Alternative operationalization of dependent variable (non-logged) | WA-D.5 |
| • Alternative format of data: User × date × one hour panel | WA-D.6 |
| • Additional controls for unobservables | |
| - Double Machine Learning (DML) | WA-D.7 |
| - Include country-level linear time trends | WA-D.7 |
| - Replace week fixed effects with daily fixed effects | WA-D.7 |
| • Replications with subsets of data | |
| - Use data from the week of the Twitch interruption only | WA-D.8 |
| - Drop data after the Twitch interruption | WA-D.8 |
| - Replicate with users accounting for 99% of playtime | WA-D.8 |
| - Replicate with users accounting for 95% of playtime | WA-D.8 |
| • Potential long-run or carryover effects of the outage | WA-D.9 |
| • Alternative modeling framework to account for zeros: Poisson model | WA-D.10 |

Note: WA denotes Web Appendix.

## 5. Conclusion

The creator economy, especially in entertainment categories like video games, often involves a fundamentally different value creation process than the standard value chain. In the standard value chain, value creation is conceptualized as a linear process where steps in the chain, ranging from material sourcing to product recycling, create value for consumers. In the livestreaming creator economy, consumers often derive value from participation in a distributed fandom community rather than from a single product. In the case of video games, consumers are members of fandoms who derive value from a "core" video game and "peripheral" products such as live-streamed content related to the game, esports competitions, or conventions featuring activities like cosplay (costume play). These are complex environments. While the value directly associated with the core product is straightforward, the peripheral but related original content produced by creators can enhance the core product's value through entertainment, education, or other forms of engagement.

A challenge in this type of digital consumption ecosystem is that since a single firm does not coordinate the core product and peripheral content, potential conflicts arise. The core product maker and the content creators share some objectives, as both benefit from a stronger fandom or consumption community. Still, content creators are also potential competitors, as they exist in a competitive space and try to maximize their followings and engagement. Potentially, ancillary content may become a competitor to the core product if consumers choose to watch creator content rather than play the core product. Further complicating the situation is that content creators can be valuable marketing partners for the core firm. When content creators specialize in categories like gaming or fashion, they can develop specialized audiences that are the primary targets for game publishers or fashion designers.

Our research focuses on this critical aspect of social media and streaming platforms, the potential benefits and conflicts involved when independent creators produce content centered on source material from fandom-oriented categories such as video games, sports, and music. In this section, we summarize our main findings and discuss our contributions to theory, practice, and directions for future research.

### 5.1. Contribution to Research and Theory

Our research extends the literature on the creator economy in several ways. First, we use linked individual-level streamer viewing and game-playing data to explore the direct, individual-level connection between viewing live-streamed content and use of an endorsed product. Granular data is a significant addition to the literature, given that most datasets available in this context are scraped at the game and streamer level (Tian et al. 2024, Huang and Morozov 2025, Li et al. 2025). The platform-level data cannot reveal how individuals sequence viewing and play, which creators they repeatedly engage with, or how a given user's play responds to a viewing shock, precisely the dimensions our analysis exploits.

In terms of co-consumption, we extend our understanding of the role of peripheral content creators by examining the underlying consumer behavior that shapes ***how*** their content complements the core material. Specifically, we investigate two plausible consumption processes: concurrent and sequential consumption of live-streamed content and video game products. We find that sequential consumption is the more dominant behavior among users. The pattern of sequential consumption matters because it suggests a feedback mechanism where viewing leads to play. This is an interesting consumer behavior finding as it suggests players enjoy the game content in multiple forms, but separately. This is meaningful, as it may enable source material publishers to establish more beneficial incentive structures for their partners. If viewing and playing were concurrent, the ideal incentive structure would emphasize the length of sessions. Given that consumption is sequential, the ideal incentive structure would emphasize conversion rates within specific time windows.

Notably, our research begins to address a significantly under-researched aspect of the creator economy: the role of user-streamer relationships. While academic research on the role of creators is not new in marketing, most studies have focused on creator-level heterogeneity (Tian et al. 2024, Li et al. 2025) or the use of livestreaming platforms from the perspective of a core product that sponsors live streamers (Jo and Lewis 2024, Huang and Morozov 2025). These efforts provide valuable insights for business practice, but our study differentiates itself by focusing on one of the most critical aspects of the creator economy: the types of creators that users repeatedly engage with. When consumers choose to watch specific creators repeatedly, relationships begin to form, and those creators may become more influential.

Given the structure of social media networks, studies of consumer-influencer relationships are a critical addition to the literature. Our empirical analyses investigate whether users' concentrated viewing toward specific types of streamers affects the relative effectiveness of live-streamed content within creator-player relationships in motivating consumption of the core product. This perspective provides a new way to understand the dynamics of the creator economy. Some content creators may offer slick productions, others may provide official information, while others may offer interactive, relationship-oriented experiences. We find that the smaller, more interactive streamers are more influential.

One interpretation is that micro streamers, who remain active community members themselves, function as community builders whose content reinforces collective participation in the core product. Mega streamers, by contrast, face stronger incentives to cultivate their own brand-centered audiences rather than strengthen the game's core community. This distinction between community-embedded and community-adjacent creators suggests a promising direction for future research on how the structural roles of different creator types within fandom ecosystems shape their downstream influence on product engagement.

### 5.2. Contribution to Managerial Practice

Informing managerial practice in the creator economy requires understanding the incentives of publishers, creators, and consumers. Understanding the interrelationships between creators, firms, and consumers highlights two critical issues. First, creators and publishers have incentives to cooperate and compete. Understanding whether and how ancillary content is a complement or substitute is critical to understanding the creator economy. Second, conceptualizing a creator economy as a consumption or fandom community is vital because it highlights the potential role of relationships between consumers and independent creators. These relationships are potentially a foundational element of these communities, and the impact of consumer-creator relationships may significantly impact source material firms’ sponsorship strategies and creators' audience management tactics.

Understanding the potential competition and opportunities for cooperation between content creators and core product consumption requires insight into the end consumers or fans. In our context, the users are fans of a video game who both play it and watch live-streamed content about it. Users play a critical role because they ultimately determine the success of both the content creators and the sponsoring firms. Their engagement with the created content drives visibility, reach, and influence, while their purchasing decisions directly impact the return on investment for sponsoring brands.

Our research offers several managerial implications for the practice of selecting external creators to endorse products. First, we find a positive effect of live-streamed content consumption on the consumption of an endorsed product. This is an important finding, as managers need to justify the marketing budget allocated to working with creators to increase product awareness. Our estimate suggests an elasticity of .308, meaning that a 10% increase in live-streamed viewing leads to approximately a 3.08% increase in gameplay time. This elasticity should help guide sponsorship rates.

Second, our research challenges the most common approach in the industry, hiring "big-name" streamers with massive audiences. Our findings indicate that when users' viewing becomes more concentrated on mega streamers, the relative effectiveness of live-streamed content decreases, revealing that attention focused on creators does not always translate into greater engagement with the endorsed product. In contrast, we find that more relationship oriented viewing of micro streamers substantially increases the positive effect of live-streamed content. However, although micro streamers drive stronger per-user responsiveness, mega streamers still generate the largest aggregate impact due to their broad reach. Decisions on collaboration should reflect the trade-off between depth of engagement and breadth of exposure, recognizing the nuances between "big names" and "niche artists."

Third, we find an intermediate result when viewers' viewing is more concentrated on the publisher's channel. The publisher's channel has an influence in between the micro and mega streamers despite being the largest single channel. This is important because it implies that investing in a firm-owned streaming channel does not adversely affect the complementarity between live-streamed content and the firm's product, unlike popular mega streamer channels.

Finally, our findings on user-level heterogeneity in viewership patterns may be valuable for many firms, as they can inform a targeting strategy to categorize customers into multiple segments. Given that the marketing budget for user engagement is often limited, understanding who is more receptive to content created by live streamers can provide valuable insights for firms seeking to promote trials of their content.

### 5.3. Limitations and Avenues for Future Research

The study is subject to limitations that suggest avenues for future research. First, our data is limited to existing players of the specific video game. This data is appropriate for our goal of understanding the link between the consumption of live-streamed content and the consumption of games. However, creators can also influence customer acquisition. Future research might examine the relationship between watching streams and new game downloads. This type of study would require greater access to user data from streaming platforms. Second, our sample is limited to players who linked their Twitch and game accounts, which is required to jointly observe viewing and play. This linkage defines the population to which our estimates apply. Since the outage is exogenous and the linking decision predates it, the sample construction does not threaten the identification of the effect. Linked players are likely more streaming-engaged than the broader player base, so our estimates are most directly informative for the segment that firms target with streamer sponsorships, where viewing behavior is the relevant margin.

Third, our study does not consider in-game purchases, even though they are a critical revenue source for freemium games. Future work could examine long-term or cumulative effects of live-streamed content on monetization. Fourth, while our identification strategy leverages a Twitch outage as a natural experiment, this disruption does not perfectly mirror firm-controlled interventions (e.g., selectively banning or promoting streamers). As with many natural experiments, the exogenous shock provides clean identification but may not align fully with managerial levers. Future work could explore settings that more closely approximate firm-driven policies. Finally, our study does not consider the unstructured data generated in livestreaming because most videos are not archived. It would be worthwhile to extract metrics on streamer-viewer interactions and investigate whether they are associated with subsequent plays and purchases.

Mega Influencers versus Niche Creators:

An Empirical Study of Streamer Influence on Endorsed Product Usage

# Web Appendix

### *A. Country-level Streaming Preferences*

Table A reports the share of users who spend the majority of their viewing time (≥70% viewing share) with firm-owned, mega, or micro streamers across nine countries from our research sample. This provides additional descriptive context on the distribution of primary streamer types at the country level. The results show that mega streamers dominate in most countries, although firm-owned and micro streamers have relatively more presence in some regions (e.g., Poland, Ukraine).

Table A: Primary Streamer (Viewing share ≥ 70%) Type by Country

| Country | N Users | Firm-owned channel | Mega streamers | Micro streamers |
|---|---|---|---|---|
| U.S. | 6,150 | 9.2% | 49.2% | 7.8% |
| Spain | 1,209 | 8.4% | 49.3% | 6.9% |
| U.K. | 1,088 | 11.2% | 49.0% | 7.1% |
| Germany | 586 | 13.8% | 35.3% | 15.7% |
| France | 335 | 17.0% | 39.1% | 9.0% |
| Poland | 246 | 21.1% | 30.1% | 19.1% |
| Romania | 236 | 28.0% | 37.3% | 3.4% |
| Ukraine | 157 | 23.6% | 17.8% | 26.8% |
| Türkiye | 126 | 11.1% | 52.4% | 4.8% |

### *B. Details on Twitch Server Down Events*

Here, we provide further details about the unexpected service outage of Twitch.tv that occurred from August 30 to 31, 2017. We present a set of anecdotal evidence to demonstrate the existence of this exogenous shock.

#### *B.1. Twitch Announcements on Social Media*

Once the website began experiencing issues with user access, Twitch's support team started investigating the problem. However, since the website was not loading properly for many users, the firm used their support team's Twitter account to inform users about the issue and provide updates on their progress. Table B below shows the timeline of the event with screenshots from their social media account. This indicates that the service interruption began around 3 PM on August 30 (US Eastern Time) and continued until at least 1:00 AM on August 31 (US Eastern Time), which is the period used for our identification.

Table B: Tweets of @TwitchSupport at Twitter

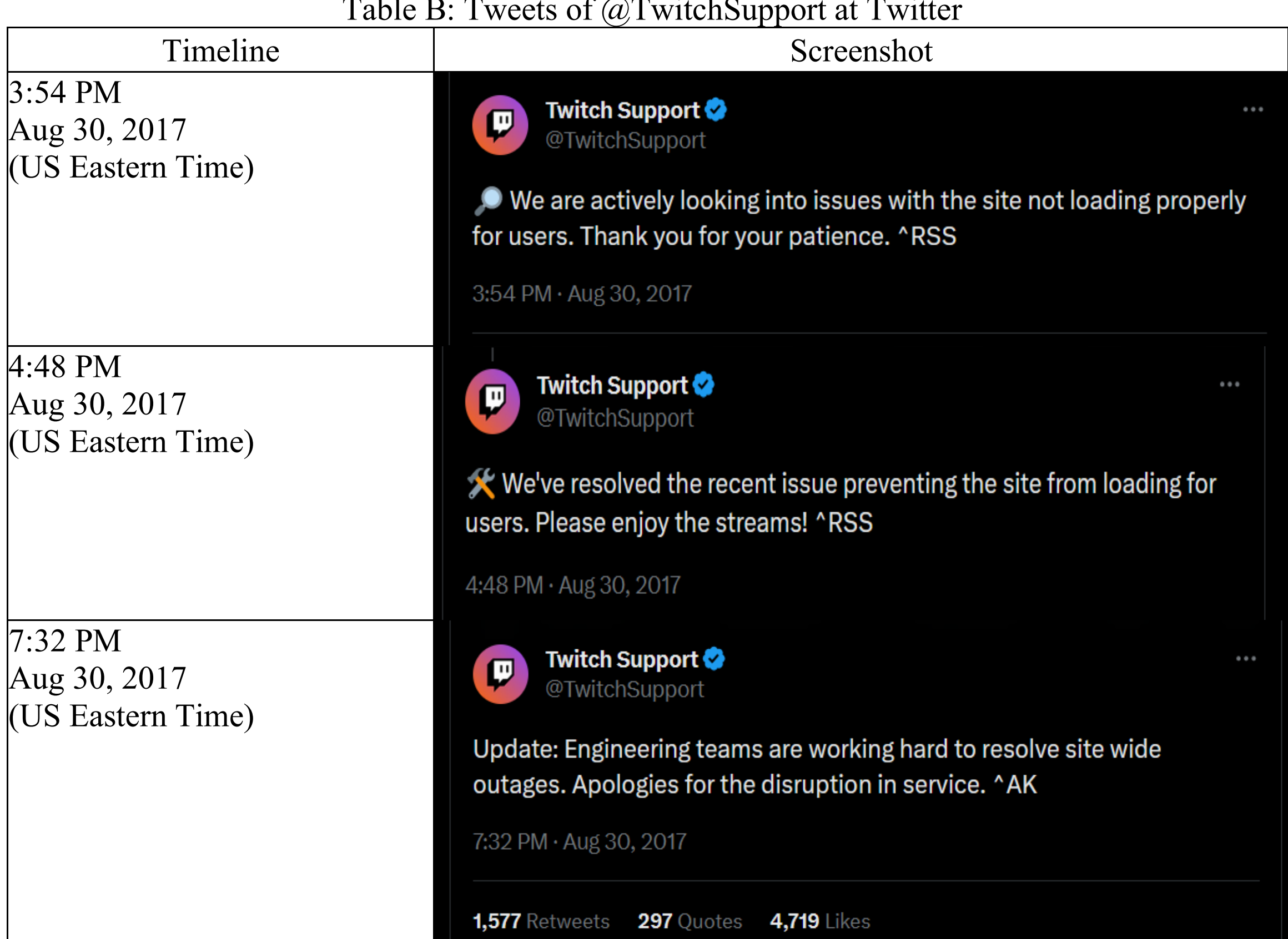

| Timeline | Screenshot |
| --- | --- |
| 3:54 PM<br>Aug 30, 2017<br>(US Eastern Time) | Twitch Support @TwitchSupport<br>🔎 We are actively looking into issues with the site not loading properly for users. Thank you for your patience. ^RSS<br>3:54 PM · Aug 30, 2017 |
| 4:48 PM<br>Aug 30, 2017<br>(US Eastern Time) | Twitch Support @TwitchSupport<br>🛠 We've resolved the recent issue preventing the site from loading for users. Please enjoy the streams! ^RSS<br>4:48 PM · Aug 30, 2017 |
| 7:32 PM<br>Aug 30, 2017<br>(US Eastern Time) | Twitch Support @TwitchSupport<br>Update: Engineering teams are working hard to resolve site wide outages. Apologies for the disruption in service. ^AK<br>7:32 PM · Aug 30, 2017<br>1,577 Retweets 297 Quotes 4,719 Likes |

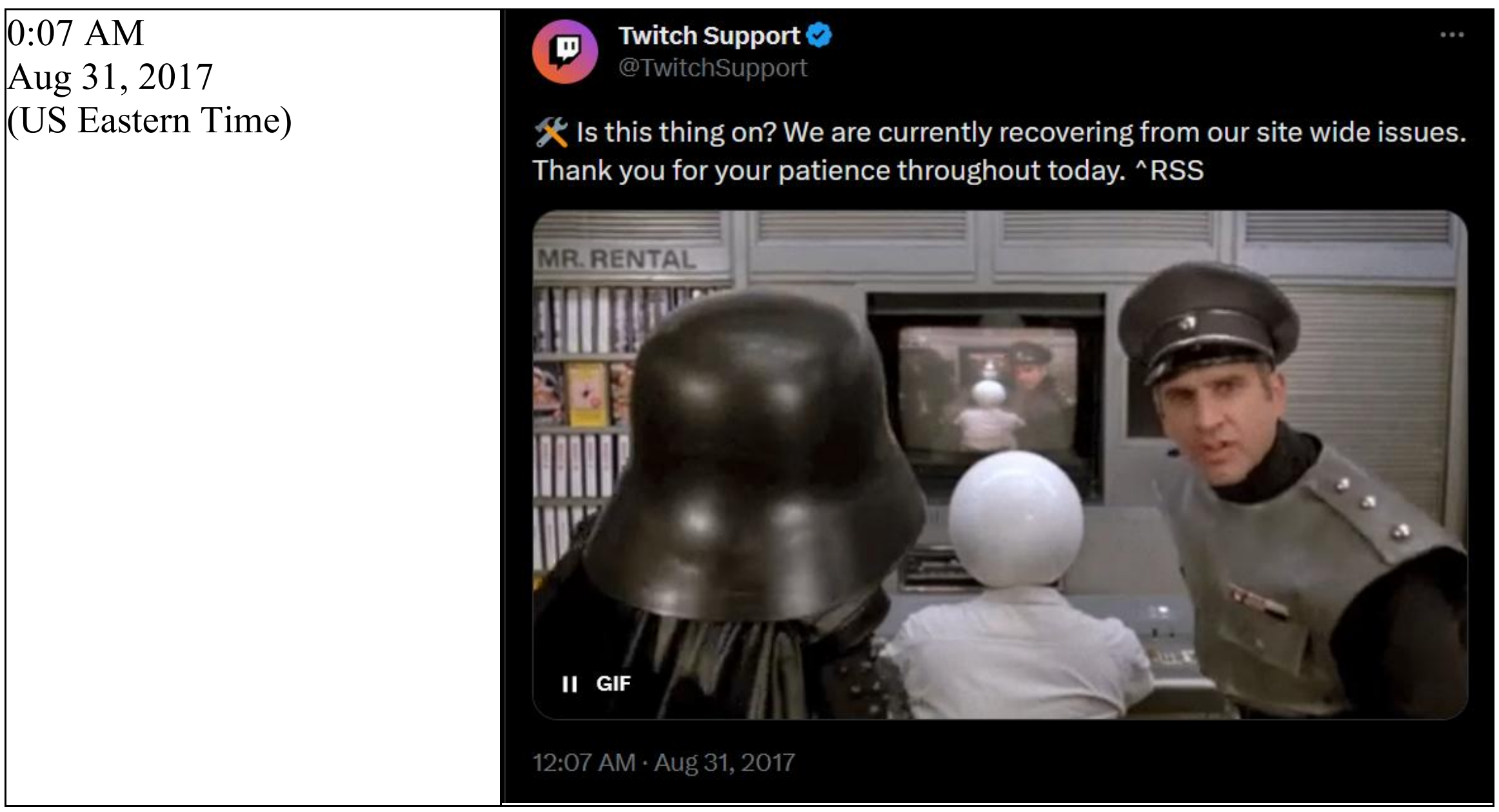

*B.2. "Twitch Down" Search Volume Measured on Google Trends*

When a favorite website does not load properly, the natural reaction is to use a search platform to find out if something is wrong with the site. If a service outage truly occurred, we would expect to see a sudden spike in web searches inquiring about the status of Twitch. To capture this, we obtained worldwide search trends for the keyword "Twitch down" for approximately two weeks before and after the outage period. As illustrated in Figure B below, there is a clear and significant surge in search volume during the two-day period (August 30-31, 2017) when the interruption occurred.

Figure B: Google Trends Score for "Twitch Down"

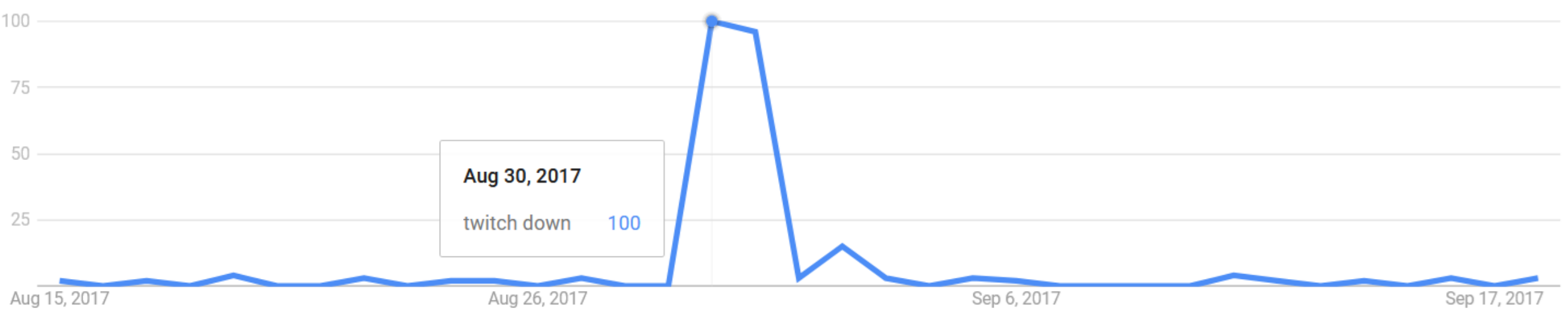

Note: The raw data for this figure can be accessed at (public url):
https://trends.google.com/trends/explore?date=2017-08-15%202017-09-18&q=twitch%20down&hl=en

*C. Supplementary Exhibits Supporting Identification*

*C1. Determination of Typical Entertainment Hours Within a Day*

In the main manuscript, we assume that people in any country typically engage with video game products from the late afternoon (4 PM) to midnight. While this assumption is reasonable, we ensure that this pattern is supported by our dataset. To validate this, we revisit the raw data and construct a country-day-hour panel dataset. The dataset includes nine countries from our sample, 63 unique days during our observation period (July 18 to September 18, 2017), and 24 unique hours within a day. For each observation, we count the number of game matches initiated at each country, day, and hour. Finally, we create a country-by-hour plot showing the average standardized number of game matches played, with 95% confidence intervals. These patterns are presented in Figure C1 below.

Figure C1: Hourly Game Usage Patterns by Country

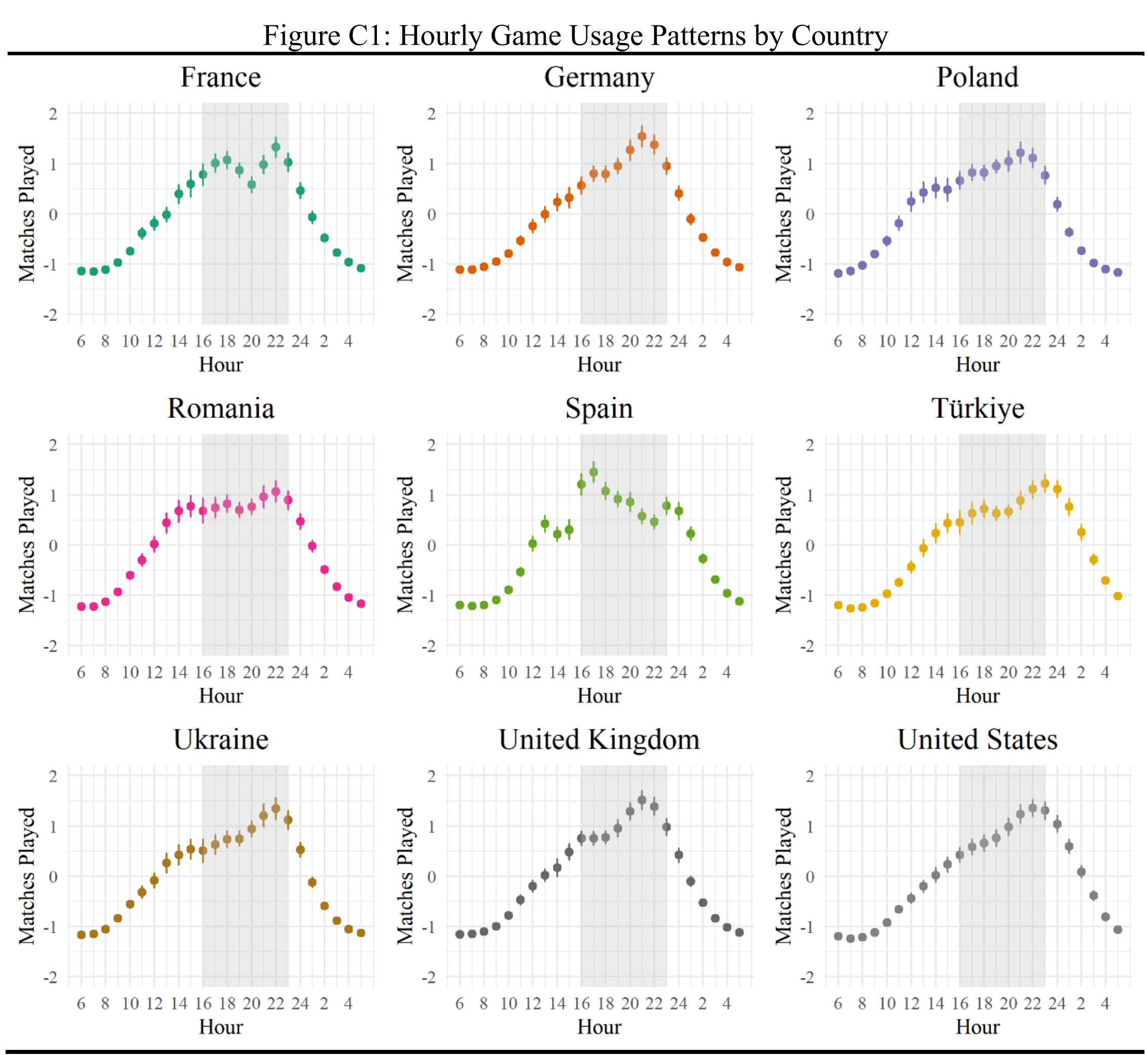

Note: Colors are available online. Dots represent standardized game matches played by country, with 95% confidence intervals shown as error bars. All gaming hours are in local time. The x-axis covers a 24-hour period, starting at 6:00 AM, with labels in two-hour intervals. The grey area highlights matches played from 4:00 PM to 11:59 PM.

Figure C1 suggests that in nearly all countries, game usage starts to increase sharply around 4 PM and begins to wind down around midnight. Therefore, we define the eight hours from 4 PM to midnight as the typical entertainment hours for players in our *4.2. Geography-Based Analysis* section. As a follow-up, we create individual-level typical entertainment hours to compute the elasticity of live-streamed content on gameplay. Further details are provided in *4.3 Effect of Live Streamed Content Viewing on Gameplays*.

### *C2. Hourly Playtime Within a Day by Country*

Figure C2 illustrates standardized hourly gameplay patterns for nine countries in our dataset. The plots show an inverted U-shaped pattern of gameplay, with activity peaking in the evening (around 6–10 PM local time) and declining toward the morning hours. The red shaded regions indicate the timing of the Twitch outage in each country. While cross-country differences exist, the general temporal pattern is consistent, reinforcing the plausibility of using time-of-day variation in our identification strategy. We observe that the outage overlapped most severely with peak evening hours in the United States, and this overlap with peak hours becomes weaker in countries located further east.

Figure C2. Hourly Playtime Within a Day by Country

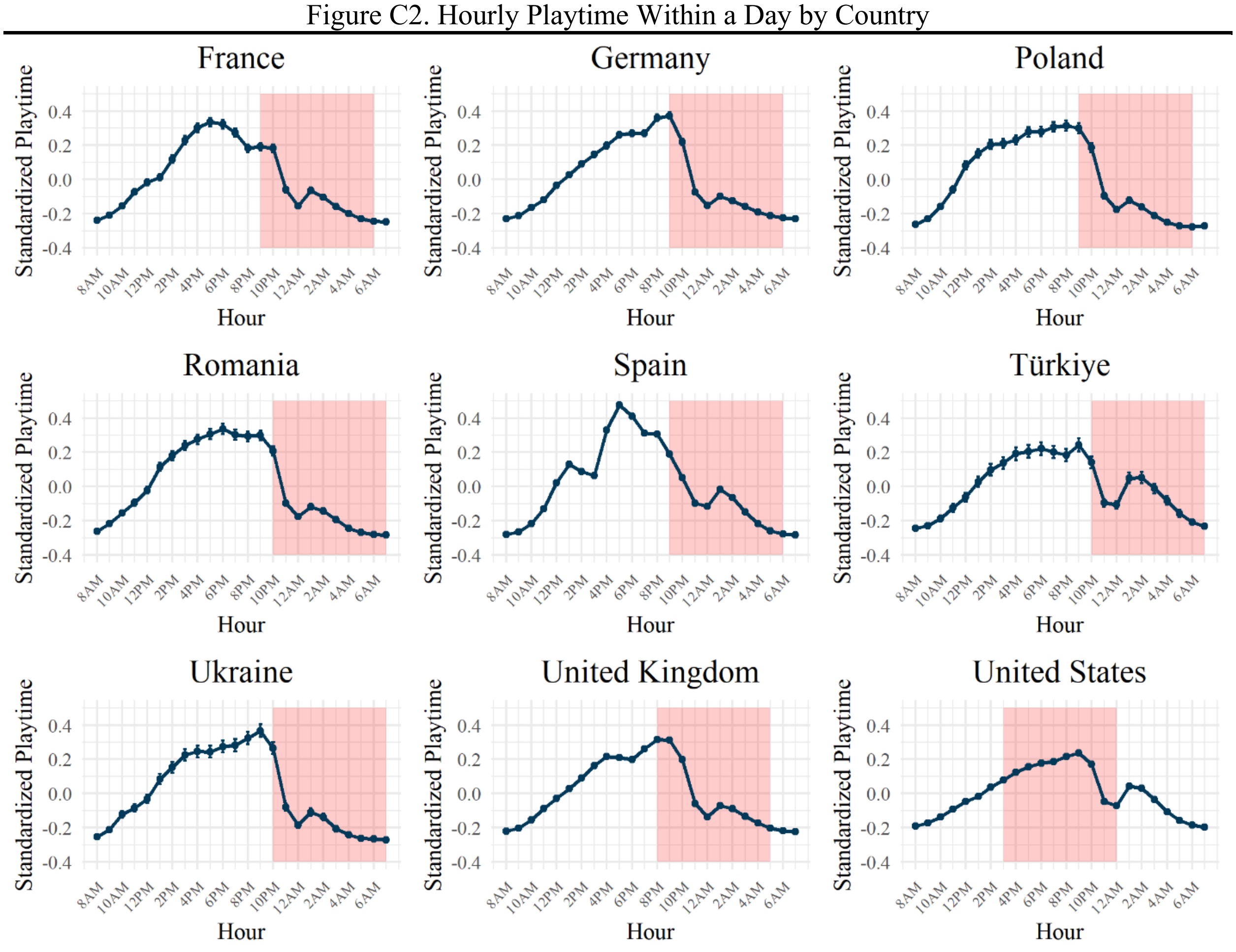

Note: The average play minutes per hour is standardized at the country level. Hours during which Twitch outages occurred are shown in red. US data is scaled to Eastern Time. Error bars denote 95% confidence intervals.

### *D. Robustness Check*

### *D.1. Robustness Checks: Alternative operationalization of "Down" dummy*

We operationalize the $\text{Down}_{c(i)th}$ dummy variable in Equation (2) by defining a three-hour block as affected if at least two of its hours were impacted by the Twitch interruption. To ensure robustness, we redefine the $\text{Down}_{c(i)th}$ dummy by assigning a value of one to three-hour blocks if **any portion** of the block overlapped with the Twitch interruption. The results of our analyses are presented in Table D1 below. We find that the main findings remain qualitatively consistent.

Table D1: Robustness Checks Using Alternative Operationalization of Down Dummy

| Description | First-stage | Second-stage: Main | Second-stage: Interaction |
|---|---|---|---|
| Dependent variable | $\ln(\text{Viewmins}_{ith}+1)$ | $\ln(\text{Playmins}_{ith}+1)$ | $\ln(\text{Playmins}_{ith}+1)$ |
| Model | (1) | (2) | (3) |
| $\ln(\text{Viewmins}_{ith}+1)$ | | .322*** | .320*** |
| | | (.035) | (.039) |
| $\ln(\text{Viewmins}_{ith}+1)\times \boldsymbol{ViewShare}_i^{\text{FirmOwned}}$ | | | -.058** |
| | | | (.018) |
| $\ln(\text{Viewmins}_{ith}+1)\times \boldsymbol{ViewShare}_i^{\text{Mega}}$ | | | -.068*** |
| | | | (.008) |
| $\ln(\text{Viewmins}_{ith}+1)\times \boldsymbol{ViewShare}_i^{\text{Micro}}$ | | | .176*** |
| | | | (.019) |
| $\text{Down}_{c(i)th}$ | -.226*** | | |
| | (.007) | | |
| $\text{Down}_{c(i)th}\times \text{AffectedHours}_i$ | -.014*** | | |
| | (.003) | | |
| $\text{Tenure}_{it}$ | -.332** | -.332* | -.333** |
| | (.112) | (.135) | (.110) |
| $\hat{\omega}_{ith}$ (control function) | | -.139*** | -.134*** |
| | | (.035) | (.039) |
| F-statistics of: | | | |
| $\text{Down}_{c(i)th}$ | 1,132.72*** | | |
| $\text{Down}_{c(i)th}\times \text{AffectedHours}_i$ | 27.84*** | | |
| Individual user FEs | Yes | Yes | Yes |
| Weekly time FEs | Yes | Yes | Yes |
| Day-of-week × Three-hour block FEs | Yes | Yes | Yes |
| R squared | .192 | .183 | .184 |
| R squared (adj.) | .189 | .180 | .181 |

Note: *p<.05; **p<.01; ***p<.001. The number of observations is 2,837,240. For the first stage, clustered standard errors at the user level were used. For the second stage, bootstrapped standard errors from 50 resamples were used. The moderators are all centered on their means.

### *D.2. Robustness Checks: Alternative time zone assumption for U.S. sample*

Although we have granular information regarding users' locations, we face challenges with the U.S. sample due to the country's diverse time zones. Currently, we assume that every U.S. user resides in the Eastern Time Zone, as this zone covers about 48% of the total population, followed by 29% in the Central Time Zone, 6.7% in the Mountain Time Zone, and 16.6% in the Pacific Time Zone (RPS Relocation 2018). However, this assumption may not always be correct and may lead to the potential misclassification of some users' time zones.

To ensure the robustness of our results and account for potential biases from such misclassification, we implement a sensitivity analysis by assuming that **30%** of U.S. users reside in the Pacific Time Zone (UTC-7). This proportion is twice the actual population share, providing a conservative adjustment to test the impact of potential time zone misclassification.

We randomly select 30% of the U.S. sample, reassign these users to the Pacific Time Zone, and adjust the relevant down dummy variable accordingly. We then re-estimate the first and second stage models. This resampling procedure is repeated 100 times to ensure that our results are not driven by spurious patterns in the data. Figure D1 below shows the distribution of the coefficient of interest from 100 replications. Overall, we successfully replicate the positive effect. The empirical 95% confidence interval in grey dashed lines do not include zero. This provides further assurance that potential misclassification of users' time zones within the U.S. sample will not seriously distort our empirical findings.

Figure D1: Estimated Elasticity When 30% of U.S. Samples Are Assumed to Be in the Pacific Time Zone

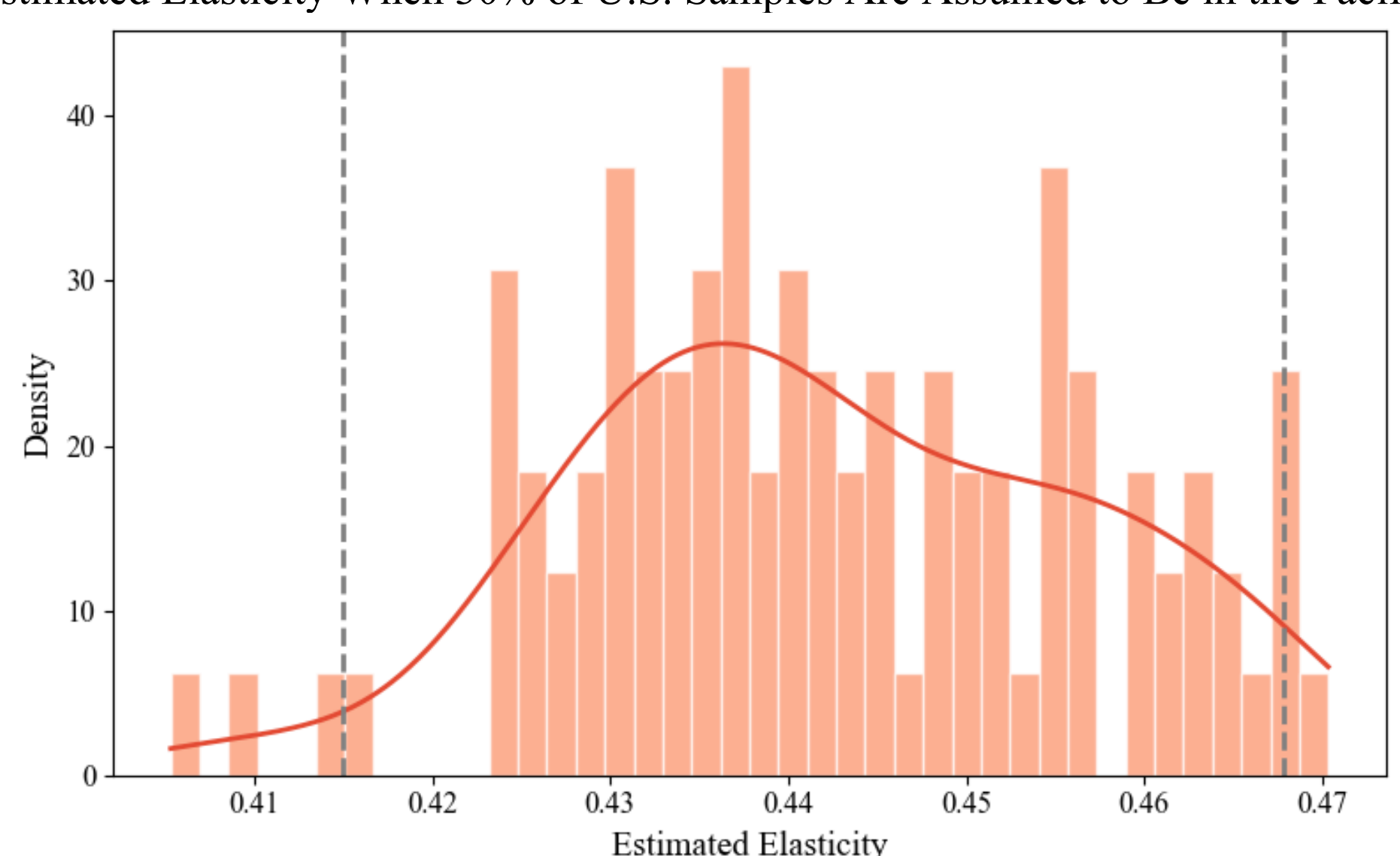

Note: Color available online. Grey dashed lines represent the 95% empirical confidence interval.

### *D.3. Robustness Checks: Alternative operationalizations of affected hours*

In the manuscript, we define the preferred gaming time within a day by identifying the top **eight** most played gaming hours based on the calibration period data. However, these eight hours may either be too few or too many to accurately capture an individual's gaming preference. To ensure the robustness of our findings, we replicate the main analyses using alternative definitions of prime gaming hours. Specifically, we compute the top **six** and top **ten** most played gaming hours within a day and calculate the new affected hours that overlap with Twitch's server down events. We then re-estimate the first-stage model, obtain the new control function, and replicate the second-stage model. The replicated second-stage results are presented below.

Table D2: Robustness Checks Using Alternative Operationalizations of Affected Hours

| $\text{AffectedHours}_i$ (instrument) defined as: | Top six hours | Top eight hours | Top ten hours |
|---|---|---|---|
| Model | (1) | (2) | (3) |
| $\ln(\text{Viewmins}_{ith} + 1)$ | .327*** | .340*** | .313*** |
| | (.043) | (.039) | (.042) |
| $\ln(\text{Viewmins}_{ith} + 1) \times \boldsymbol{ViewShare}_i^{\text{FirmOwned}}$ | -.058** | -.058*** | -.058** |
| | (.019) | (.017) | (.018) |
| $\ln(\text{Viewmins}_{ith} + 1) \times \boldsymbol{ViewShare}_i^{\text{Mega}}$ | -.068*** | -.068*** | -.068*** |
| | (.009) | (.007) | (.008) |
| $\ln(\text{Viewmins}_{ith} + 1) \times \boldsymbol{ViewShare}_i^{\text{Micro}}$ | .176*** | .176*** | .176*** |
| | (.019) | (.016) | (.016) |
| $\text{Tenure}_{it}$ | -.330* | -.326 | -.335** |
| | (.135) | (.167) | (.111) |
| $\widehat{\omega}_{ith}$ (control function) | -.142*** | -.155*** | -.127** |
| | (.043) | (.040) | (.042) |
| R squared | .183 | .184 | .184 |
| R squared (adj.) | .180 | .181 | .181 |

Note: *p<.05; **p<.01; ***p<.001. The sample consists of 2,837,240 observations. All models include user, week, and day-of-week × three-hour block fixed effects. Bootstrapped standard errors with 50 resamples are used and they are in parentheses. The moderators are all centered on their means.

***D.4. Robustness Checks: Alternative operationalization for category viewership variable***

In the main manuscript, we construct the category viewership variable by first defining meaningful user–streamer ties as repeated visits during the calibration period, and we have already demonstrated robustness using different thresholds: four, eight, and twelve visits.

However, one may argue that absolute viewing time matters more when defining meaningful ties. To ensure robustness to this alternative rationale, we adopt additional criteria based on total viewing time. Specifically, we define a user as having a meaningful relationship with a streamer if their cumulative viewing time during the calibration period exceeds **60, 90, or 120 minutes**. The operationalization of the category viewership variable remains the same: the fraction of time devoted to all streamers with meaningful ties across the three streamer types.

As reported in Table D3, we find that all patterns from the main analysis are replicated even when using alternative definitions of meaningful ties.

Table D3: Robustness to Alternative Definitions of the Meaningful Streamer-User Ties

| Meaningful tie is defined as: | *Watching **60** minutes or more* during the calibration period | *Watching **90** minutes or more* during the calibration period | *Watching **120** minutes or more* during the calibration period |
|---|---|---|---|
| Model | (1) | (2) | (3) |
| $\ln(\text{Viewmins}_{ith} + 1)$ | .340$^{***}$ | .341$^{***}$ | .341$^{***}$ |
| | (.040) | (.040) | (.041) |
| $\ln(\text{Viewmins}_{ith} + 1) \times \boldsymbol{ViewShare}_i^{\text{FirmOwned}}$ | -.144$^{***}$ | -.124$^{***}$ | -.114$^{***}$ |
| | (.017) | (.014) | (.012) |
| $\ln(\text{Viewmins}_{ith} + 1) \times \boldsymbol{ViewShare}_i^{\text{Mega}}$ | -.144$^{***}$ | -.127$^{***}$ | -.117$^{***}$ |
| | (.016) | (.011) | (.011) |
| $\ln(\text{Viewmins}_{ith} + 1) \times \boldsymbol{ViewShare}_i^{\text{Micro}}$ | .066$^{***}$ | .085$^{***}$ | .101$^{***}$ |
| | (.017) | (.019) | (.018) |
| $\text{Tenure}_{it}$ | -.321$^{*}$ | -.316$^{*}$ | -.314$^{*}$ |
| | (.137) | (.142) | (.148) |
| $\widehat{\omega}_{ith}$ (control function) | -.154$^{***}$ | -.153$^{***}$ | -.154$^{***}$ |
| | (.041) | (.040) | (.041) |
| R squared | .185 | .185 | .185 |
| R squared (adj.) | .182 | .182 | .182 |

Note: *p<.05; **p<.01; ***p<.001. The sample consists of 2,837,240 observations. All models include user, week, and day-of-week × three-hour block fixed effects. Bootstrapped standard errors with 50 resamples are used and they are in parentheses. The moderators are all centered on their means.

### *D.5. Robustness Checks: Alternative definition of dependent variable*

We test the robustness of our findings by using the non-log-scaled dependent variable (play minutes) and confirm that the main results remain qualitatively consistent.

Table D4: Replication with Alternative Definition of Dependent Variable (non-logged y)

| Variables | Est. | S.E. |
|---|---|---|
| $\ln(\text{Viewmins}_{ith} + 1)$ | 6.062*** | .818 |
| $\ln(\text{Viewmins}_{ith} + 1) \times \boldsymbol{ViewShare}_{i}^{\text{FirmOwned}}$ | -1.268*** | .362 |
| $\ln(\text{Viewmins}_{ith} + 1) \times \boldsymbol{ViewShare}_{i}^{\text{Mega}}$ | -1.406*** | .189 |
| $\ln(\text{Viewmins}_{ith} + 1) \times \boldsymbol{ViewShare}_{i}^{\text{Micro}}$ | 4.439*** | .515 |
| $\text{Tenure}_{it}$ | -6.876** | 2.445 |
| $\widehat{\omega}_{ith}$ (control function) | -2.980*** | .825 |
| R squared | .152 | |
| R squared (adj.) | .149 | |

Note: *p<.05; **p<.01; ***p<.001. The sample consists of 2,837,240 observations. All models include user, week, and day-of-week × three-hour block fixed effects. Bootstrapped standard errors with 50 resamples are used and they are in parentheses. The moderators are all centered on their means.

### *D.6. Robustness Checks: Alternative panel scale (one hour instead of three hours)*

We test the robustness of our findings by rescaling the data from the current three-hour panel to a one-hour panel at the user-by-date level. For each hour, we use both log-scaled gameplay minutes and the ratio of gameplay minutes to the full 60 minutes. The results remain qualitatively consistent.

Table D5: Replication Using One-Hour Panel Data

| Dependent variable | $\log(playmins_{ith} + 1)$ | | $\frac{playmins_{ith}}{60}$ | |
|---|---|---|---|---|
| Variables | Est. | S.E. | Est. | S.E. |
| $\ln(\text{Viewmins}_{ith} + 1)$ | $.348^{***}$ | .044 | $.071^{***}$ | .009 |
| $\ln(\text{Viewmins}_{ith} + 1) \times \boldsymbol{ViewShare}_i^{\text{FirmOwned}}$ | $-.036^{**}$ | .014 | $-.008^{**}$ | .003 |
| $\ln(\text{Viewmins}_{ith} + 1) \times \boldsymbol{ViewShare}_i^{\text{Mega}}$ | $-.033^{***}$ | .006 | $-.007^{***}$ | .001 |
| $\ln(\text{Viewmins}_{ith} + 1) \times \boldsymbol{ViewShare}_i^{\text{Micro}}$ | $.078^{***}$ | .013 | $.016^{***}$ | .003 |
| $\text{Tenure}_{it}$ | $-.184^{*}$ | .085 | $-.037^{*}$ | .019 |
| $\widehat{\omega}_{ith}$ (control function) | $-.257^{***}$ | .044 | $-.053^{***}$ | .009 |
| R squared | .097 | | .089 | |
| R squared (adj.) | .096 | | .088 | |

Note: *p<.05; **p<.01; ***p<.001. The sample consists of 8,511,720 observations. All models include user, week, and day-of-week × one-hour block fixed effects. Bootstrapped standard errors with 50 resamples are used and they are in parentheses. The moderators are all centered on their means.

***D.7. Robustness Checks: Additional controls for unobservables***

***D.7.1. Double Machine Learning***

While our original model includes high-dimensional fixed effects, some unobserved confounding factors may still remain—for example, specific time blocks within certain countries. To address this, we apply the Double Machine Learning (DML) framework to more flexibly account for unobservables (Chernozhukov et al. 2018). Specifically, we implement the Partial Linear Instrumental Variables (PLIV) approach within the DML-based empirical framework. This method aligns well with our original empirical strategy, as it addresses both (1) the endogeneity of our key regressor (live-streaming viewing minutes) and (2) the presence of potentially complex and interactive fixed effects or covariate structures (Chernozhukov et al. 2018).

***Setup***. The variable structure shares the same structure as in the main analysis. The outcome variable is log-scaled game play minutes after adding one. The endogenous regressor is log viewing minutes after adding one. The instruments are an indicator for the outage block and its interaction with user-specific affected hours based on primary gaming hours. Controls include country, week, three-hour block, and day-of-week dummies, and log-scaled account tenure. We use the identical sample window and variable definitions as in the main model.

For estimating the nuisance components, we use Gradient Boosting Regressors. This method efficiently captures nonlinear relationships and complex interactions among covariates, making it well suited for flexibly modeling high-dimensional nuisance functions in DML. We set n_estimators = 100, max_depth = 3, learning_rate = 0.1, subsample = 1.0, random_state = 1017. All categorical variables are one-hot encoded (with the first level omitted to avoid multicollinearity).

***Estimation***. We use DoubleMLPLIV class from DoubleML Python library with K = 5 folds and cross-fitting enabled. Inference is based on the DML asymptotic variance with 200 bootstrap repetitions. All estimates are based on cross-fitted residuals, ensuring orthogonalization between the treatment and nuisance components.

***Results***. The DML-based causal estimate is .297 with a 95% confidence interval of [.175, .420], indicating a statistically significant and directionally consistent effect (p < .001), as reported in Figure D2. The estimate is slightly smaller than the elasticity from the fixed effects regression in Table 6 ($\beta$ = .343; p < .001), likely because the DML framework more flexibly absorbs high-dimensional unobserved heterogeneity. Nevertheless, the two estimates largely overlap in their 95% confidence intervals, confirming the robustness of our findings.

Figure D2. Comparison of Causal Effect Estimates: Fixed Effects vs. DML-PLIV

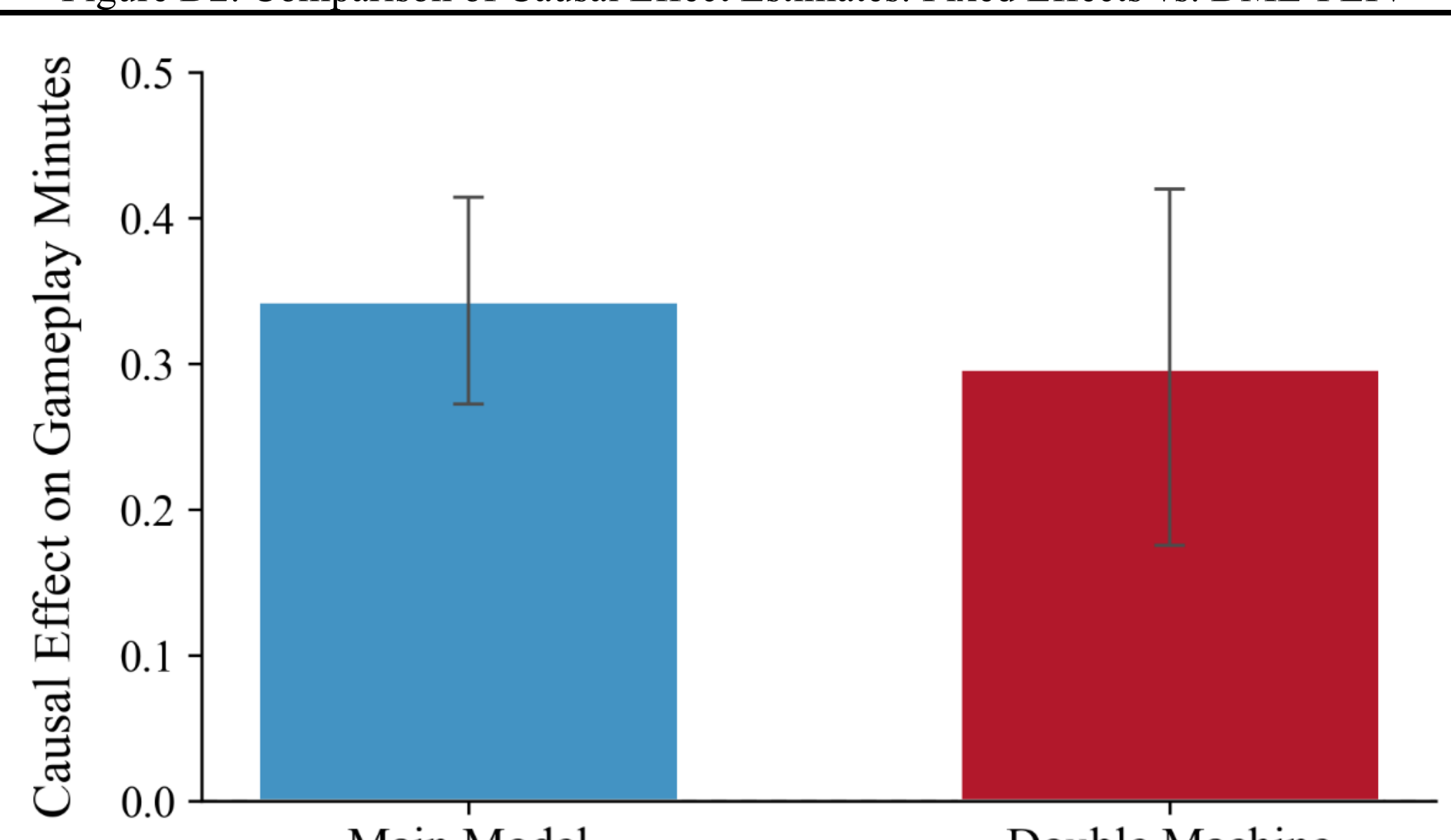

Note: Both bars represent the estimated elasticity of livestreaming viewing minutes on gameplay minutes. Error bars indicate 95% confidence intervals.

### *D.7.2. Additional Controls*

We test the robustness of our findings by adding additional controls to address potential unobservable factors. Table D6 reports the results of these robustness checks. First, we assume a linear time trend in each country and replicate the second-stage moderation model by including a country-specific daily linear time trend (model 1). Second, we replace the weekly fixed effects with daily fixed effects (model 2). Finally, we include additional day-of-week fixed effects to account for specific temporal trends within each day of the week. Across all robustness checks, we continue to observe results that are qualitatively consistent with our main analyses.

Table D6: Robustness Checks—Additional Controls for Unobservables

| Additional Controls | Country-level linear time trends | Daily time FEs |
|---|---|---|
| Model | (1) | (2) |
| $\ln(\text{Viewmins}_{ith} + 1)$ | .341*** | .434*** |
| | (.046) | (.043) |
| $\ln(\text{Viewmins}_{ith} + 1) \times \boldsymbol{ViewShare}_i^{\text{FirmOwned}}$ | -.058*** | -.058*** |
| | (.017) | (.017) |
| $\ln(\text{Viewmins}_{ith} + 1) \times \boldsymbol{ViewShare}_i^{\text{Mega}}$ | -.068*** | -.068*** |
| | (.006) | (.008) |
| $\ln(\text{Viewmins}_{ith} + 1) \times \boldsymbol{ViewShare}_i^{\text{Micro}}$ | .176*** | .175*** |
| | (.017) | (.020) |
| $\text{Tenure}_{it}$ | -.291* | -.291** |
| | (.130) | (.104) |
| $\hat{\omega}_{ith}$ (control function) | -.156*** | -.249*** |
| | (.045) | (.043) |
| Individual user FEs | Included | Included |
| Time FEs | Weekly time FEs | Daily time FEs |
| Day-of-week × 3-hour block FEs | Included | Included |
| Additional controls | Country-level linear time trends | |
| R squared | .184 | .184 |
| R squared (adj.) | .181 | .182 |

Note: *p<.05; **p<.01; ***p<.001. The number of observations is 2,837,240. Bootstrapped standard errors with 50 resamples are used and are in parentheses. The moderators are all centered on their means.

### *D.8. Robustness Checks: Replications with subsets of data*

Another set of analyses replicates our results using subsets of the data, with the findings reported in Table D7. First, we restrict the sample to the week of the Twitch interruption, since the instrument-induced variation is concentrated in this period. This ensures that our findings are not driven by observations without instrument-induced variation (column 1). Second, we replicate the analysis after excluding all observations following the outage, to verify that our findings are not influenced by potential long-run spillover effects (column 2). Third, because the focal game is a freemium title, some users are highly loyal while others play only rarely. To focus on economically meaningful behavior, we re-estimate the models using the top users who account for 99% and 95% of total game time (columns 3 and 4). Across all these subsamples, the core findings of our analyses remain robust.

Table D7: Robustness Checks—Replications with Subsets of Data

| Subset | Only use the week of Twitch outage | Drop observations after outage | Samples accounting for 99% of total usage | Samples accounting for 95% of total usage |
|---|---|---|---|---|
| Model | (1) | (2) | (2) | (3) |
| $\ln(\text{Viewmins}_{ith} + 1)$ | .381*** | .537*** | .356*** | .364*** |
| | (.062) | (.041) | (.048) | (.043) |
| $\ln(\text{Viewmins}_{ith} + 1) \times \boldsymbol{ViewShare}_i^{\text{FirmOwned}}$ | -.093*** | -.051* | -.062** | -.077*** |
| | (.027) | (.022) | (.020) | (.019) |
| $\ln(\text{Viewmins}_{ith} + 1) \times \boldsymbol{ViewShare}_i^{\text{Mega}}$ | -.074*** | -.064*** | -.070*** | -.066*** |
| | (.012) | (.009) | (.010) | (.011) |
| $\ln(\text{Viewmins}_{ith} + 1) \times \boldsymbol{ViewShare}_i^{\text{Micro}}$ | .148*** | .190*** | .183*** | .191*** |
| | (.026) | (.022) | (.016) | (.022) |
| $\text{Tenure}_{it}$ | 2.938*** | -.817** | -.427** | -.507** |
| | (.876) | (.303) | (.150) | (.181) |
| $\widehat{\omega}_{ith}$ (control function) | -.207** | -.354*** | -.160*** | -.157*** |
| | (.064) | (.042) | (.048) | (.044) |
| R squared | .230 | .204 | .178 | .175 |
| R squared (adj.) | .216 | .199 | .175 | .172 |
| Observations | 567,448 | 1,378,088 | 2,526,160 | 2,141,720 |

Note: *p<.05; **p<.01; ***p<.001. All models include user, week, and day-of-week × three-hour block fixed effects. Bootstrapped standard errors with 50 resamples are used and they are in parentheses. The moderators are all centered on their means.

### *D.9. Robustness Checks: Potential long-run or carryover effects of the outage*

In our main analysis, we relied on the unexpected Twitch outage as a source of identification. However, a potential concern is that the outage may have created carryover effects beyond its immediate timing. If the outage had lasting effects on game usage or stream content viewing behavior, it could potentially bias our identification strategy, which relies on non-outage days as part of the counterfactual. Conversely, if there were no substantial lasting effects, this would, in fact, strengthen our approach by allowing us to leverage the outage event in identifying the treatment effect.

To address this concern, we conducted a new analysis on whether there was any effect after the outage. Specifically, we created a dummy variable ($\text{AfterDown}_{ith}$) that takes the value of one (1) if it falls within (1) the 24 hours after the Twitch outage and (2) the next seven days after the outage. We restricted the sample to the pre-outage period and the first seven days following the end of the outage to compare post-outage patterns with pre-outage baselines, since the post-outage window may capture any lingering effects of the disruption. These measures capture any differences in behavior following the outage over one-day and seven-day periods, which should provide sufficient time to observe emerging patterns.

Table D8: Lasting Effect of Twitch Outage

| Dependent variable | $\ln(\text{Playmins}_{ith} + 1)$ | | $\ln(\text{Viewmins}_{ith} + 1)$ | |
|---|---|---|---|---|
| Model | (1) | (2) | (3) | (4) |
| $\text{Down}_{ith}$ | -.077*** | -.072*** | -.201*** | -.207*** |
| | (.010) | (.010) | (.007) | (.007) |
| $\text{AfterDown}_{ith}$(Next 1 day) | -.008 | | .018*** | |
| | (.006) | | (.005) | |
| $\text{AfterDown}_{ith}$(Next 7 days) | | .006 | | -.004 |
| | | (.005) | | (.004) |
| $\text{Tenure}_{it}$ | -.625** | -.634** | -.630*** | -.625*** |
| | (.204) | (.205) | (.157) | (.159) |
| R squared | .176 | .176 | .202 | .202 |
| R squared (Adj.) | .172 | .172 | .198 | .198 |

Note: *p<.05; **p<.01; ***p<.001. The sample consists of 1,872,957 observations, with standard errors clustered at the player level (in parentheses). User-, week-, day-of-week by three-hour block fixed effects are included.

We report the results of our analysis in Table D8. In columns (1) and (2), we observe that although game usage significantly dropped during the outage (consistent with the main manuscript), there is no significant effect in the periods immediately following the outage, whether within one day or seven days. This clearly indicates that game usage returns to its usual level once everything returns to normal.

In columns (3) and (4), we replicate the same analysis for live-streaming content viewing behavior. We find that on the day immediately after the outage, there is a slight and significant increase in users' viewing minutes, with an approximate 1.8% increase (column 3). However, this behavior flattens over time and eventually dissipates within a week, as the effect observed after seven days (column 4) is insignificant. This suggests that although users exhibit slight compensatory behavior on the day following the outage, this effect diminishes over time, disappearing within a week.

***D.10. Robustness Checks: Alternative modeling framework to account for zeros: Poisson model***

While our dataset offers a great deal of granularity, the cost is that the dependent variable is zero in a large share of observations (about 84% of cells, Table 4). This is not unexpected, as users do not play the game in every three-hour block within a day. However, the dominance of zeros, combined with the log transformation of the dependent variable, may raise empirical concerns. Specifically, the $\ln(\text{Playmins}_{ith} + 1)$ transformation is sensitive to the units of the dependent variable, and with a large mass at zero the implied elasticity can depend on this arbitrary scaling (Chen and Roth 2024).

To address this, we re-estimate our heterogeneity specification (Equation 4) using Poisson pseudo-maximum likelihood (PPML). PPML models gameplay minutes directly in levels, requires no log transformation, and accommodates the large share of zeros naturally, remaining consistent under correct specification of the conditional mean without further distributional assumptions (Santos Silva and Tenreyro 2006). We retain the same control function correction, fixed effects, and mean-centered category viewership shares as in our main model.

Table D9 reports the results. The pattern across streamer types remains unchanged: the interaction with micro-streamer viewership share is positive and significant, while those with firm-owned and mega-streamer shares are negative. The pairwise contrasts confirm our central finding that the viewing-to-play response is strongest for micro-concentrated viewers.

Table D9: Poisson Fixed Effect Model Result

| Variables | Estimate | S.E. |
|---|---|---|
| $\ln(\text{Viewmins}_{ith} + 1)$ | .191*** | .054 |
| $\ln(\text{Viewmins}_{ith} + 1) \times \boldsymbol{ViewShare}_i^{FirmOwned}$ | -.077*** | .007 |
| $\ln(\text{Viewmins}_{ith} + 1) \times \boldsymbol{ViewShare}_i^{Mega}$ | -.018*** | .003 |
| $\ln(\text{Viewmins}_{ith} + 1) \times \boldsymbol{ViewShare}_i^{Micro}$ | .173*** | .005 |
| $\text{Tenure}_{it}$ | -.685*** | .095 |
| $\widehat{\omega}_{ith}$ (control function) | -.032 | .054 |

Note: *p<.05; **p<.01; ***p<.001. The sample consists of 2,837,240 observations. All models include user, week, and day-of-week × three-hour block fixed effects. Bootstrapped standard errors with 50 resamples are used and they are in parentheses. The moderators are all centered on their means.

### *E. External Validity: Generalizability Analysis*

A potential concern regarding our main findings is whether they are merely artifacts of a single video game dataset. While we believe this is not likely—given that the focal game shares many key features common to typical games on live streaming platforms, such as real-time multiplayer mechanics, frequent content updates, and active engagement with the streaming community—it is important to assess the generalizability of our results. To this end, we replicate our analysis using daily data on player counts and Twitch viewership for 268 other video game titles.

Specifically, we collected game-level data from *SteamDB* (https://steamdb.info/), an independent web database that provides detailed information on the Steam platform and its games. *SteamDB* covers nearly all popular video games globally and offers daily data on active users and Twitch viewership for each game title. Because the platform prohibits automated scraping, we contacted the data manager and received approval to scrape and use the data for academic research.

- We selected our sample of games using the following process:
- Scrape a list of 1,000 popular video games.
- Visit each of the $k$th game title pages and collect daily active user counts and Twitch viewer data.
- Repeat this process until reaching the 1,000th game.
- Trim the data to match the five-week observation period used in our main analysis.
    - Games that do not have valid data during the observation period are dropped.

As a result, we retained 268 unique game titles with valid observations within the specified timeframe. The resulting game-by-date panel dataset consists of 9,233 observations. The dataset includes many globally popular video game titles such as *Grand Theft Auto V, Counter-Strike: Global Offensive, Dota 2, PlayerUnknown's Battlegrounds (PUBG), Tom Clancy's Rainbow Six Siege, Left 4 Dead 2, The Witcher 3: Wild Hunt, ARK: Survival Evolved, and Sid Meier's Civilization VI*.

Finally, we supplemented this dataset with Twitch viewership distribution data for August 2017 by visiting each game title's page on *SullyGnome* (https://sullygnome.com/), an independent platform that archives detailed Twitch analytics. We extracted viewership distribution data for the period from August 1 to August 29, 2017 (the day before the Twitch outage). SullyGnome reports the percentage of viewership held by the 1st, 2nd, 3rd, 4th, 5th, 6th–10th, 11th–25th, 26th–50th, 51st–100th, and 101st–250th most popular creator channels. Using this data, we computed **the average proportion of viewership held by the top five creators** as a measure of the extent to which a game's Twitch viewership is concentrated among a few top streamers. Our rationale is that, in the absence of user-level category viewership measures, it is reasonable to assume that category viewership to mega streamers is most pronounced in games where a higher share of viewership is concentrated among the top five creators. However, we acknowledge that this is an imperfect proxy rather than a direct measure of the "category viewership" construct examined in our main analysis. We recognize this as a limitation of analyses that rely on publicly available aggregate data.

We first regressed the log-scaled Twitch viewer counts (a potentially endogenous variable) on the Twitch outage time dummies used in the main analysis. While the main analysis leveraged both between-country and within-day variation in Twitch outages, this supplementary analysis was limited to using daily time indicators to capture the effect. Nonetheless, following the same logic behind our instrumental variable strategy, we incorporated the **relative popularity of each game title in the United States**, given that the outage primarily affected users on the western side of the Atlantic. To measure this, we computed the Google Trends search score for each game title during August 2017 and used the score recorded for the U.S. as a proxy for the game's popularity in the U.S. relative to the rest of the world. We then interacted the outage dummy with this U.S. popularity measure and used the resulting term to construct the control function. The corresponding results are reported in Table E1.

Table E1. First-stage Regression

| *Variable* | *Estimate* | *S.E.* |
|---|---|---|
| $Down_t$ | -.1196* | .0579 |
| $Down_t \times US\ Popularity_i$ | -.3995* | .1807 |
| Game Fixed Effects | Included | |
| Weekly Fixed Effects | Included | |
| R squared | .8200 | |
| R squared (adj.) | .8140 | |

Note: *p<.05. The number of observations is 9,233. Standard errors are clustered at game level.

Both coefficients in the first-stage model indicate a negative and statistically significant effect, suggesting that Twitch viewer numbers declined on the dates of the outage, with a more pronounced drop for games that were popular in the U.S. These findings are consistent with our earlier results in the manuscript, supporting the view that the Twitch outage significantly disrupted viewing behavior among American players.

Next, we generate the residuals from the first-stage regression and use them as a control function. In the second-stage model, we estimate the effect of log-scaled Twitch viewer counts on the log-scaled active player counts for each game. To examine the heterogeneity in creator composition—consistent with our main analyses—we divide the games into three groups based on the average proportion of viewership captured by the top five creator channels. We adopt this subgrouping approach, rather than interacting the moderator directly, because creator concentration is defined at the game level in these supplemental data, making subgroup comparisons a more transparent way to assess cross-game heterogeneity.

Figure E1 below plots the coefficients of log-scaled Twitch viewers for three subgroups. Overall, all estimates are positive and directionally consistent with our main analysis, suggesting a complementary relationship between live-streamed content and the promoted products. However, the figure also shows that as the proportion of viewership accounted for by the top five major creators increases (from the Low group to the High group), the complementary effect becomes weaker—reinforcing the core pattern observed in our main findings. In unreported analyses, we also examined whether similar patterns emerge when using alternative moderators such as the viewership concentration of the top two streamers and a viewership diversification index (Herfindahl-Hirschman Index). These analyses similarly reveal heterogeneous effects, consistent with the notion that concentration of attention among top creators decreases the strength of complementarity.

Figure E1. Heterogenous Complementary by Viewership Concentration

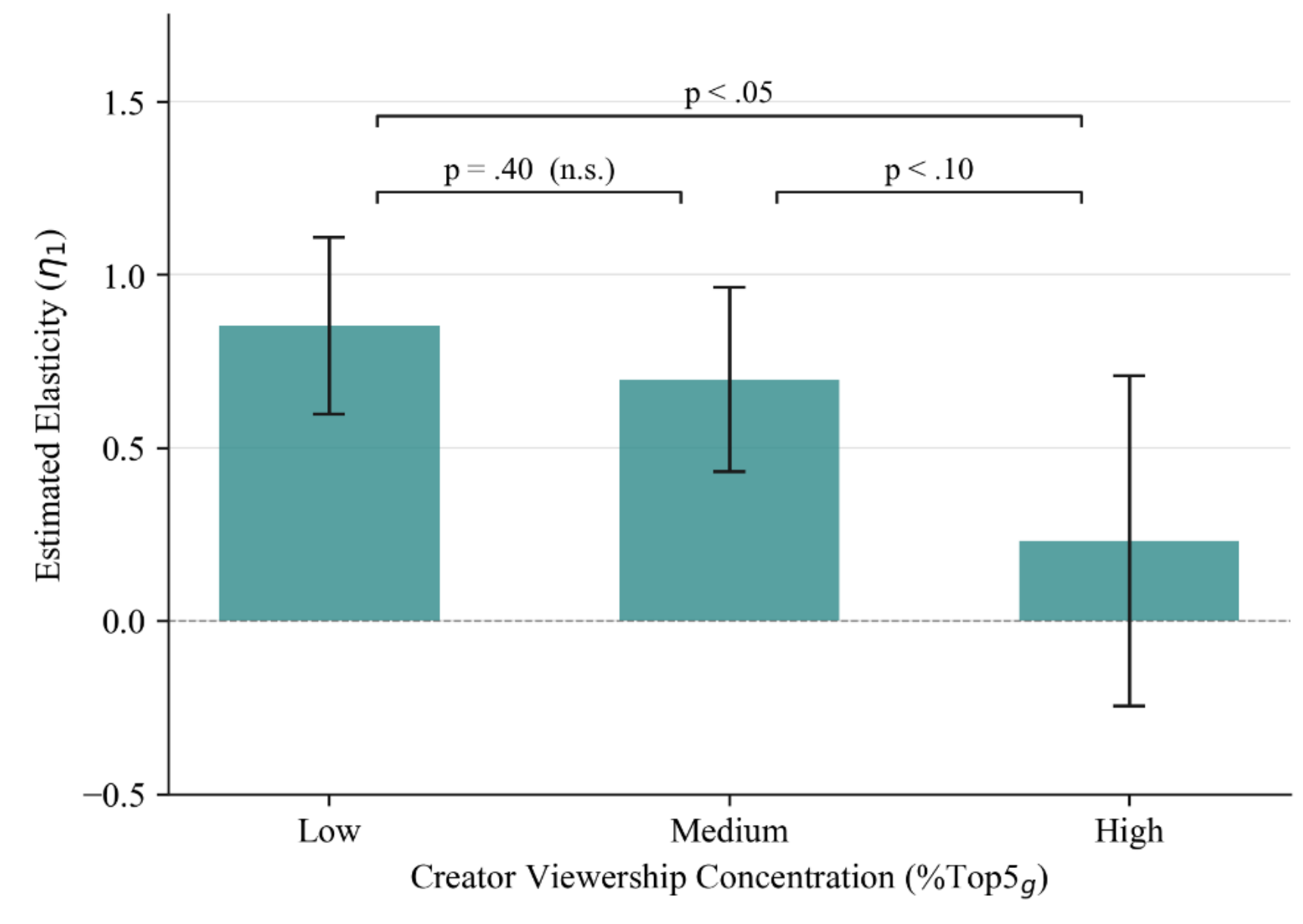

Note: Color is available online. Error bars denote 95% confidence intervals for the elasticity estimates.

We also replicate the findings using an interaction-based regression on the pooled data, as described in equation WA.E. Table E2 reports the estimated results of this pooled regression. The difference between the low and medium groups is not statistically significant. However, the elasticity estimate decreases significantly among games with high concentration on top streamers (the high group). Point estimates computed from the regression (bottom of Table E2) reconfirm this pattern.

$$\begin{aligned} \ln(Players_{gt}) = \eta_1 \ln(Viewers_{gt}) & \\ + \eta_2 \ln(Viewers_{gt}) \times \boldsymbol{Low}{:}\, \%Top5_g & \\ + \eta_3 \ln(Viewers_{gt}) \times \boldsymbol{Low}{:}\, \%Top5_g & \\ + \alpha_g + \tau_w + \hat{\varphi}_{gt} + \varepsilon_{gt} & \end{aligned} \qquad \text{WA.E}$$

Table E2. Pooled Regression with Interaction

| Variable | Estimate | S.E. |
|---|---|---|
| $\ln(Viewers_{gt})$ (baseline) | $.697^{***}$ | .136 |
| $\ln(Viewers_{gt}) \times \boldsymbol{Low}{:}\, \%Top5_g$ | .155 | .188 |
| $\ln(Viewers_{gt}) \times \boldsymbol{High}{:}\, \%Top5_g$ | $-.466^{+}$ | .278 |
| Elasticity Estimate from the Pooled Regression | | |
| Game with Low $\%Top5_g$ | $.853^{***}$ | .130 |
| Game with Medium $\%Top5_g$ | $.697^{***}$ | .136 |
| Game with High $\%Top5_g$ | .232 | .242 |
| Difference between High and Low | $-.621^{*}$ | .275 |

Note: +p<.10, *p<.05, **p<.01, ***p<.001. Standard errors are clustered at the game level. The medium tercile is the omitted baseline; implied elasticities equal the corresponding subgroup estimates. N = 9,233 (268 games).

Table E3 reports the elasticity estimates using a median split instead of terciles. Complementarity still weakens as livestreaming viewership becomes more concentrated on the top five streamers.

Table E3. Alternative Split (Median)

| Median Split Group | Elasticity Estimate | S.E. | N |
|---|---|---|---|
| Bottom 50% | $.782^{***}$ | .103 | 4,600 |
| Top 50% | $.387^{+}$ | .200 | 4,633 |

The supplementary analysis offers two key contributions to our main findings. First, it replicates the observed complementarity between live-streamed content consumption and consumption of the source material (i.e., the video game). Second, and more importantly, it shows that the moderation of this

complementarity—depending on the concentration of viewership among top mega creators—is consistently observed across a broader set of video game titles. This directly supports the generalizability of our findings, indicating that the observed patterns are not confined to the idiosyncrasies of a single game but are evident across the wider gaming ecosystem. Second, this analysis underscores the value of our granular, individual-level dataset. While limited to one game, it enables us to trace a direct link between live-streamed content consumption and the consumption of the promoted game itself. Furthermore, it allows us to characterize user-level heterogeneity in viewing behavior by examining the relative attention paid to mega- versus micro-creators and how this moderates the complementarity between streaming and gameplay. Given the consistency in results across both datasets, we believe our main findings are robust and provide actionable insights for both platform managers and game publishers.

### *References for Web Appendix*